%% file: paper.tex
\documentclass[sigplan, screen, nonacm]{acmart}
\usepackage[subtle]{savetrees}

\usepackage{epsfig}
\usepackage{balance}
\usepackage{graphicx} 
\usepackage{subcaption}
\usepackage{tabularx}

\usepackage{xspace}
\usepackage{listings}
\usepackage{verbatim}
\usepackage{booktabs}
\usepackage{colortbl}

\usepackage{nicefrac}
\usepackage{siunitx}

\usepackage{amsmath}
\usepackage{xfrac}

\usepackage{color}
\definecolor{darkred}{rgb}{0.7,0,0}
\definecolor{darkgreen}{rgb}{0,0.5,0}
\hypersetup{colorlinks=true,
        linkcolor=darkred,
        citecolor=darkgreen}

\usepackage{fancyvrb}
\usepackage{relsize}

\usepackage{algorithm,algorithmicx}
\usepackage[noend]{algpseudocode}
\usepackage[normalem]{ulem}

\usepackage[small,compact]{titlesec}
\usepackage{wrapfig}
\usepackage{blindtext}
\usepackage{makecell}
\titlespacing*{\section}{0pt}{4pt}{4pt}
\titlespacing*{\subsection}{0pt}{4pt}{4pt}
\titlespacing*{\subsubsection}{0pt}{0pt}{0pt}

\input{defs}

\begin{document}

\title{\LARGE Scalable Tail Latency Estimation for Data Center Networks}
\author{Kevin Zhao}
\affiliation{
  \country{}
  \institution{\textit{University of Washington}}
}
\author{Prateesh Goyal}
\affiliation{
  \country{}
  \institution{\textit{Microsoft Research}}
}
\author{Mohammad Alizadeh}
\affiliation{
  \country{}
  \institution{\textit{MIT CSAIL}}
}
\author{Thomas E. Anderson}
\affiliation{
  \country{}
  \institution{\textit{University of Washington}}
}
\renewcommand{\shortauthors}{X.et al.}

\begin{abstract}
  \input{abstract.tex}
\end{abstract}

\maketitle
\pagestyle{plain}
\input{intro.tex}
\input{overview.tex}

\input{methods.tex}
\input{evaluation.tex}

\input{conclusion.tex}
\def\bibfont{\normalfont}

\bibliographystyle{abbrv}
\bibliography{refs.bib}

\appendix
\input{appendix.tex}

\end{document}

%% file: defs.tex
\newcommand{\eg}{{e.g.,} }

\newcommand{\ie}{{i.e.}, }
\newcommand{\etal}{{\em et al.}\xspace}

\newcommand{\Fig}[1]{Fig.~\ref{fig:#1}\xspace}
\newcommand{\Tab}[1]{Table~\ref{tab:#1}\xspace}
\newcommand{\Sec}[1]{$\S$\ref{s:#1}\xspace}

\newcommand{\Alg}[1]{Alg.~\ref{alg:#1}\xspace}

\newcommand{\kz}[1]{{\textcolor{purple}{KZ: {#1}}}}

\newcommand{\revv}[2] {#2}
\newcommand{\rev}[1] {#1}

\newcommand{\cut}[1]{}

\newcommand{\Para}[1]{\vspace{4pt}\noindent{\bf #1}\hspace{6pt}}

\newcommand{\sys}{\texttt{Parsimon}\xspace}

\newcommand*\Let[2]{\State #1 $\gets$ #2}
\algrenewcommand\algorithmicrequire{\textbf{Precondition:}}

\def\compactify{\itemsep=0pt \topsep=0pt \partopsep=0pt \parsep=0pt}
\let\latexusecounter=\usecounter
\newenvironment{CompactItemize}
{\def\usecounter{\compactify\leftmargin=13pt\latexusecounter}
  \begin{itemize}}
    {\end{itemize}\let\usecounter=\latexusecounter}

%% file: abstract.tex
In this paper, we consider how to provide fast estimates of flow-level tail
latency performance for very large scale data center networks.  Network tail
latency is often a crucial metric for cloud application performance that can
be affected by a wide variety of factors, including network load, inter-rack
traffic skew, traffic burstiness, flow size distributions, oversubscription,
and topology asymmetry. Network simulators such as ns-3 and OMNeT++ can provide
accurate answers, but are very hard to parallelize, taking hours or days to
answer what if questions for a single configuration at even moderate scale.
Recent work with MimicNet has shown how to use machine learning to improve
simulation performance, but at a cost of including a long training step per
configuration, and with assumptions about workload and topology uniformity that
typically do not hold in practice.

We address this gap by developing a set of techniques to provide fast
performance estimates for large scale networks with general traffic matrices
and topologies. A key step is to decompose the problem into a large number of
parallel independent single-link simulations; we carefully combine these
link-level simulations to produce accurate estimates of end-to-end flow level
performance distributions for the entire network. Like MimicNet, we exploit
symmetry where possible to gain additional speedups, but without relying on
machine learning, so there is no training delay. On a large-scale network where
ns-3 takes 11 to 27 hours to simulate five seconds of network behavior,
our techniques run in one to two minutes with accuracy within 9\% 
for tail flow completion times. 


%% file: intro.tex
\section{Introduction}

Counterfactual simulation---to answer ``what if'' questions about the interaction
of network protocols, workloads, topology, and switch behavior---has long been
used by both researchers and practitioners as a way of quantifying the effect of design
options and operational parameters~\cite{sim97,ns-3,omnet,opnet,dctcp,dcqcn,homa,hpcc}.
As production data center networks have scaled up in bandwidth and
scaled out in size~\cite{fb-fabric,jupiter}, however, network simulation has failed to keep pace.  Although
there is ample parallelism at a physical level in large scale data center networks, it has been
difficult to realize significant speedup with packet-level network simulation~\cite{parsim94,genesis}.
As packets flow through the network, the scheduling decisions at each switch affect the behavior
of every flow traversing that switch, and therefore the scheduling
decisions at every downstream switch, and---with congestion control---future flow behavior,
in a cascading web of very fine-grained interaction.
In our own experiments using ns-3~\cite{ns-3}, for example,
simulating a 384-rack, 6,144-host network on
\revv{
  a standard multicore server
}{
  a single thread of a modern desktop CPU
}
took 11 to 27 hours of wall-clock time to advance five seconds of simulated time.
\revv{}{While parallel techniques for discrete event simulation
  exist~\cite{pdes}, recent work has demonstrated their limited efficacy for
  speeding up simulations of highly interconnected data center
  networks~\cite{mimicnet}.
}
As a result, packet-level network simulation today is mostly used for small
scale studies.

\revv{
  In recent work, MimicNet proposes an alternative approach, using machine
  learning to construct a model of how different parts of the network affect
  other parts of the network.
  Although promising, MimicNet has significant limitations. Because it uses
  packet-level simulation to train the model, MimicNet can take several hours
  to train on a new workload or network configuration, limiting its utility
  for exploring a wide design space. Further, since MimicNet scales by
  replicating a model trained on a portion of the network, it is
  limited to failure-free uniform fat trees with uniform traffic among equally sized clusters.
  Of course, production networks typically run a diversity of workloads
  and suffer frequent partial or grey failures that result in
  asymmetric topologies and performance~\cite{corropt}.
  Sensitivity of network performance during failure episodes is important to
  network operators.
  Another operational use case is to
  predict the impact of planned partial network outages and incremental network
  upgrades on application performance.
}{
  The need for faster network simulation has spawned recent efforts to use
  machine learning to model how different parts of the network affect each
  other~\cite{mimicnet,deepqueuenet}.
  While promising, these approaches have several limitations.
  MimicNet requires hours-long retraining for new workloads and network
  configurations, and it only accelerates simulations of uniform fat trees
  with uniform traffic among equally-sized clusters of
  machines~\cite{mimicnet}. 
  DeepQueueNet relaxes some of MimicNet's restrictions but does not model
  congestion control, which can be a first-order determiner of
  performance~\cite{deepqueuenet}.
  %
}

This paper aims to address this gap, to develop techniques for fast \rev{approximate}
simulation of large scale networks with arbitrary workloads and topologies.
Our work involves no training step, aiming to produce near-real time results even at scale. 
\rev{In addition to reducing the cost of evaluating new protocols, 
another goal is to provide real-time decision support for network operators, such as
warnings of SLO violations if links fail~\cite{f10,mogul-nines}, advice on task placement
of communication-intensive jobs~\cite{borgomega}, and predicting the
  performance impact of planned partial network outages and upgrades~\cite{andromeda,patchpanel}.
  }

A key observation is that we could achieve high degrees of parallelism if we could somehow
disentangle the interactions between switch queues, allowing us to study the behavior of the traffic
on each link in isolation. Of course, switch queues are not in reality completely disentangled.
The packets for any particular flow experience a very specific set of conditions at each switch,
and those conditions are affected by the presence of upstream bottlenecks which
can smooth packet arrivals for competing flows at downstream switches.  The congestion response
for a flow depends on the combination of conditions at every switch along the path.

However, large scale data center networks are typically managed with the goal of delivering consistent
high performance to applications. While congestion events do occur, they are often chaotic rather
than persistent, popping up and then disappearing in different spots due to the inherent
burstiness and flow size distribution of applications, rather than due to some long-term
mismatch between demand and capacity in some portion of the network~\rev{\cite{microbursts}}. Further, we are often
interested in
\revv{
  aggregate
}{
  \emph{aggregate}
}
behavior, such as the frequency of poor flow performance, rather than
\revv{
  accurately modeling
}{}
the behavior of each individual packet or flow.

\revv{
  Thus,
}{
  To model aggregate behavior,
}
our hypothesis is that we can approximate the distribution of end-to-end flow performance
for a particular workload running on a large scale network by modeling
the frequency and magnitude of local congestion events at each link along
\revv{
  the path.
}{
  individual paths.
}
A long flow will of course experience multiple congestion events during its lifetime,
but most of these will occur at different points along the path {\em at different times}.
Modeling the effect of simultaneous congestion events, and the response of the congestion
algorithm to multiple simultaneous bottlenecks, is second order.

Our hypothesis is related to the concept of product-form solutions in queuing theory. For certain classes of queueing networks (e.g., Jackson~\cite{jackson1957networks} and BCMP networks~\cite{baskett1975open}), the equilibrium distribution of queue lengths can be written in product form, i.e., the state of an individual queue is only dependent on the traffic it receives
and not on the state of the rest of the network. These results generally require specific assumptions about job arrival processes (e.g., Poisson), service-time distributions (e.g., Exponential), and queueing/routing disciplines (e.g., FIFO or processor-sharing queues), and there has been much theoretical work on identifying classes of queueing networks that admit product-form solutions~\cite{kelly1976networks}. Although data center networks do not strictly conform to these conditions and the dynamics of each individual queue can be quite complex (e.g., due to
congestion control), our hypothesis is that product-form solutions are approximately true in most realistic settings, and therefore
we can analyze individual queues in isolation and combine the results to approximate end-to-end network behavior.

\begin{figure}[t]
  \centering
  \includegraphics[width=\linewidth]{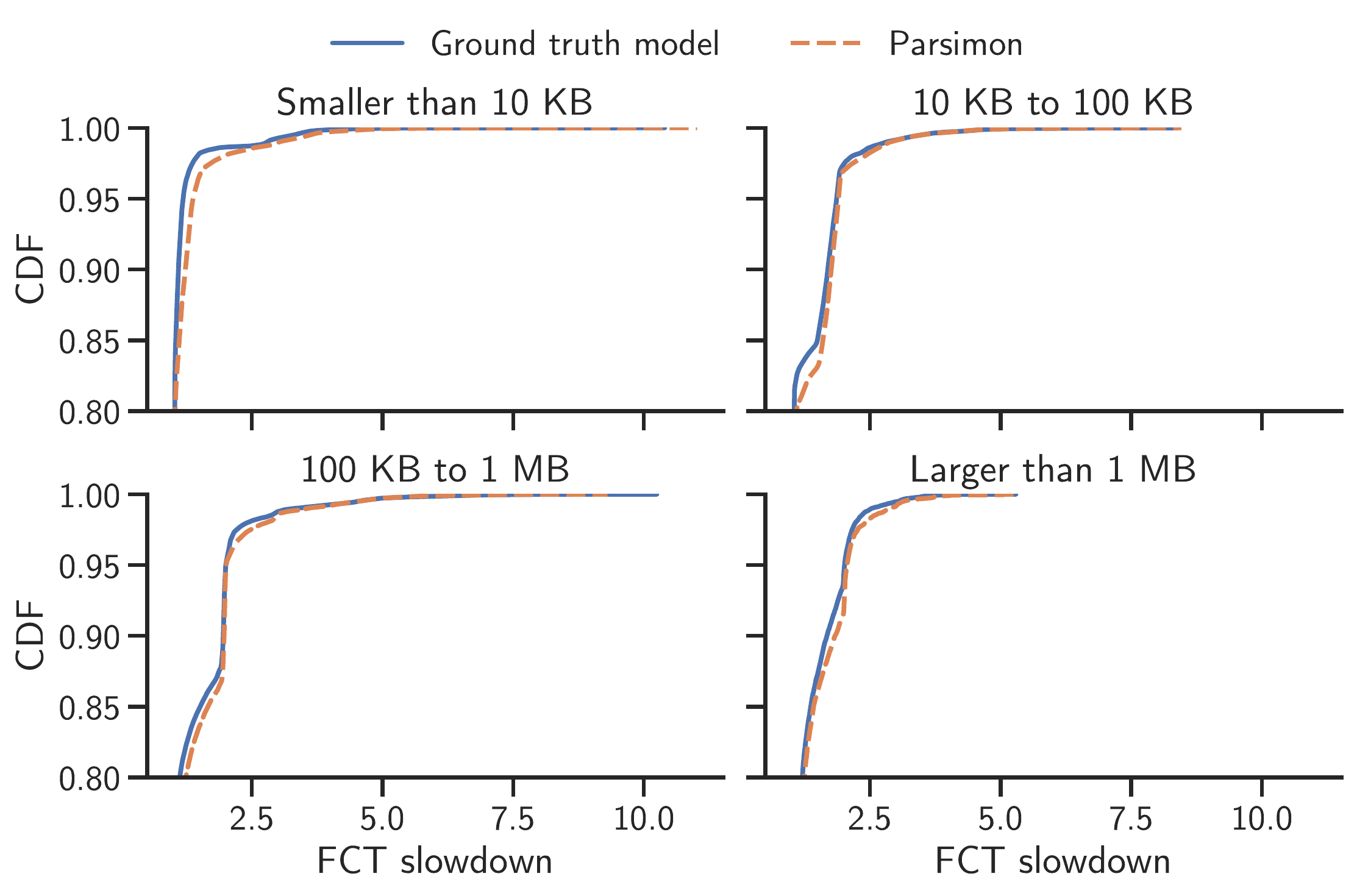}
  \vspace{-7mm}
  \caption{\small
    CDF of ns-3 versus \sys for flow completion time (FCT) slowdown across
    multiple flow size ranges, zoomed into the tail.
    While ns-3 took nearly 11 hours to produce these results, \sys took one
    minute and 19 seconds, end-to-end.
    Results were taken on a 6,144-host topology with an industry traffic
    matrix, 2-to-1 oversubscription, and bursty traffic.
  }
  \vspace{-4mm}
  \label{fig:banner}
\end{figure}

We built \sys to directly test this hypothesis.  First, we deconstruct the network topology
into a large number of simple and fast simulations where each can be run entirely in parallel by
a single hyperthread.
Each simulation aims to collect the distribution of delays that flows of a particular size
would experience through a single link, assuming that the rest of the network
is benign. We then combine these simulated delay distributions to produce predictions of
the end-to-end delay distribution, again for flows of a given size.
At each step, we make conservative assumptions for how delays should be computed and combined.
\revv{and thus our work provides a bound on the distribution of flow performance.}{}
In many settings, researchers and operators are interested in \revv{guaranteeing that tail behavior is}{keeping tail behavior}
well-managed, making a conservative assumption more appropriate than an optimistic one.
Finally, \sys clusters links with common traffic characteristics, eliminating much of the overhead
of simulating parallel links in the core of the network as well as edge links used by replicated
or parallel applications, further improving simulation performance.

Because validation against detailed packet-level simulation at scale is so expensive,
we \revv{limit our study to}{focus our study on} a single widely used transport protocol, DCTCP~\cite{dctcp}, with FIFO queues
with ECN packet marking at each switch~\cite{ecn}. 
We also focus on queue dynamics rather than packet loss; most data center networks are provisioned
and engineered for extremely low packet loss~\cite{fb-network,jupiter}.
We note that these assumptions are not fundamental to our 
\revv{approach, but validation}{approach. We show \sys generalizes to two other transport 
protocols, DCQCN~\cite{dcqcn} and the delay-based TIMELY~\cite{timely}. Validation} of other transport
protocols~\cite{hull,homa,hpcc,swift}, switch queueing disciplines~\cite{wfq,spfifo,homa,bfc}, and packet loss
remains future work. 
We note that modern data center transport layer protocols \revv{like DCTCP}{} are adept at quickly adapting to the presence
and absence of congestion, and so we caution our results may not extend to older transport protocols
where convergence time is a large factor. 

\rev{
Parsimon speeds up simulations by reasoning about links independently, which
enables massive parallelization, but at a cost in accuracy.
As we will see in \Sec{error-sources}, anything that creates standing
congestion both at the core and at the edge, or when cross traffic is correlated across multiple hops, 
will result in less accurate
estimates.
While our methods are designed to favor overestimating rather than
underestimating tail latencies, this property is only evaluated experimentally
(\Sec{eval}).
In general there is no formal guarantee, since factors like congestion control
can in theory behave in arbitrary ways that render less appropriate the
approximation of considering links independently.
%
We assume that we can simulate for long enough for the network to reach equilibrium;
studies of short term transient behavior should not use our approach. We do not provide predictions at
the level of an individual flow, but we are able to show that \sys is accurate for sub-classes of traffic
for mixed workloads.
We do not attempt to model end host scheduling delay of packet
processing, even though that may have a large impact on network performance~\cite{tailtale,swift}; we leave 
addressing that
to future work.}

\revv{}{
  To assess accuracy, we compare distributions of flow completion time (FCT)
  slowdown, defined as the observed FCT divided by the best achievable FCT on
  an unloaded network, and we say a flow is complete when all of its bytes
  have been delivered to its destination.
}
\Fig{banner} shows a sample of our results for the 6,144 host network mentioned
above, running a published industry traffic matrix~\cite{fb-network} and flow
size distribution~\cite{homa}, and with standard settings for burstiness and
over-provisioning.  We describe the details of this and other experiments later
in the paper.
Depicted are
\revv{
  distributions of flow completion time (FCT) slowdown (the ratio
  to the time the flow would take on an otherwise unloaded network),
}{
  FCT slowdown distributions
}
binned by flow size.
While ns-3 took nearly 11 hours on this configuration, \sys was able to match flow-size specific performance of ns-3
in 79 seconds (a 492 times speedup) on a single 32-way multicore server with an error
of 9\% at the 99th percentile.
Given a small cluster of simulation servers, we estimate a completion time of 21 seconds using our approach.

In our evaluation, we scan the parameter space to identify circumstances where our approximations are
less accurate.  
Link clustering improves performance but hurts accuracy somewhat; this tradeoff can be avoided by using more simulation cores.
Without clustering, accuracy suffers when there is high utilization of links in the core (above 50\%),
there are high levels of oversubscription, and a large fraction of network traffic
is due to flows that finish within a single round trip.
Generally, a combination of factors is required for poor accuracy.
In 85\% of the configurations we test, the error relative to ns-3 is under
10\%.
\revv{
  We note that MimicNet reports a mean error rate of 5\%. However, they
  test configurations with 100 Mbps links instead of the 10 Gbps and 40 Gbps
  links we assume, with the result that much less of their traffic is due to
  flows that finish within a single round trip.
}{}

\rev{\sys source code and evaluation scripts are publicly available at \url{https://github.com/netiken}.}

%% file: overview.tex
\section{\sys Overview} \label{s:overview}

\begin{figure*}[t]
  \centering
  \includegraphics[width=\linewidth]{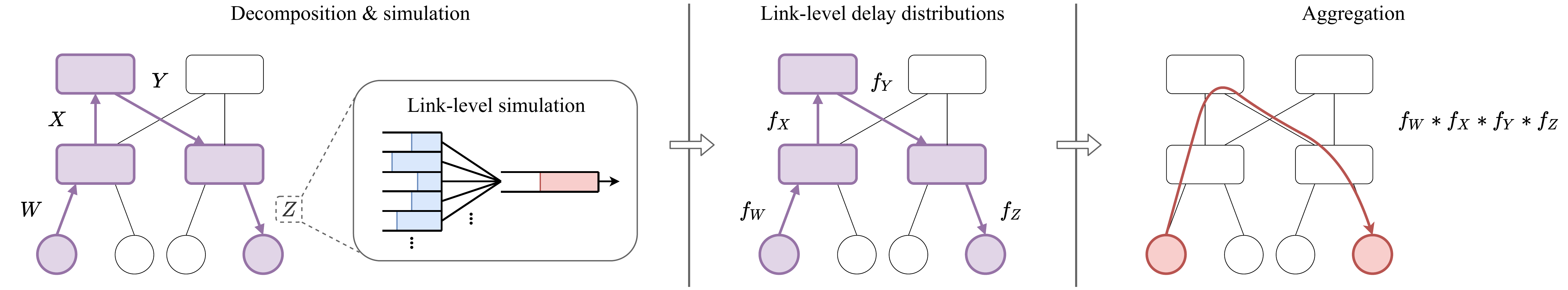}
  \vspace{-8mm}
  \caption{\small
    Overview of \sys. 
    First, for any path, \sys estimates the contribution of each
    component link to delays in flow completion times, represented as a
    probability distribution.
    \sys then combines delays along the path using Monte Carlo
    simulation (see \Sec{key-methods}).
    Further, for added performance, link-level simulations are
    optimized and redundant simulations (due to \eg ECMP or symmetries
    in workload patterns) are pruned (see \Sec{other-methods}).
  }
  \label{fig:overview}
  \vspace{-1mm}
\end{figure*}

\begin{figure}[t]
  \centering
  \includegraphics[width=0.95\linewidth]{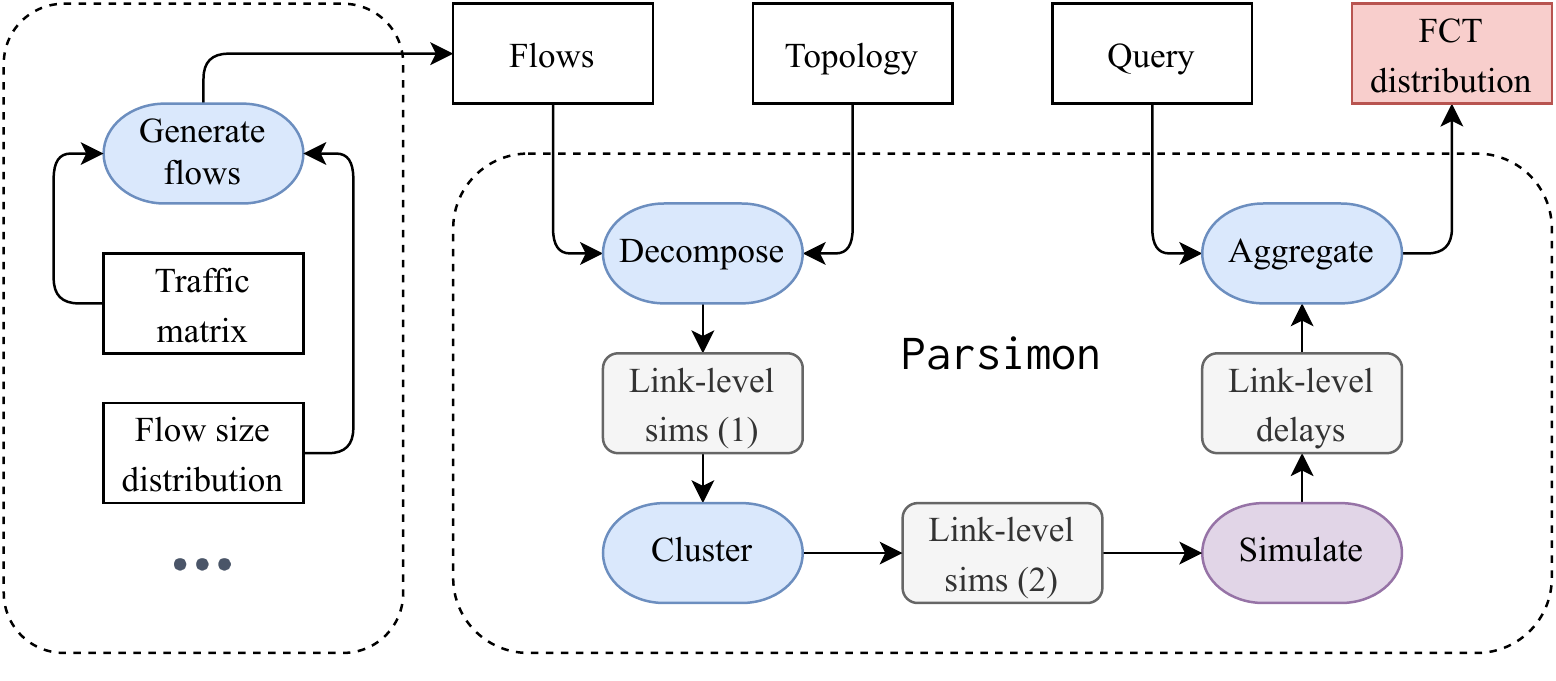}
  \vspace{-4mm}
  \caption{\small
    An illustration of \sys's workflow.
    All inputs and outputs are shown in the top row.
    Rectangular boxes are inputs and outputs, rounded boxes are
    intermediate artifacts, and ovals are \sys's actions.
  }
  \vspace{-4mm}
  \label{fig:workflow}
\end{figure}

This paper describes a set of methods to quickly and scalably estimate
\revv{
  flow performance
}{
  distributions of flow performance
}
in data center networks.
These techniques are implemented in a prototype called \sys, designed to
provide the following:
\begin{CompactItemize}
  \item {\bf Fast, scalable estimates.} We aim to supply
  estimates two to three orders of magnitude faster than full-fidelity simulation.
  Given enough cores, execution time should remain bounded regardless of network size.
  \item {\bf Tight latency bounds, including tail performance.}
  Our approximations bias slightly towards overestimation, but still provide
  close estimates even for the 95th or 99th percentile of the distribution
  for a given flow length.
  \item {\bf Minimal restrictions on topology and workload.} Our methods
  are largely independent of both topology and workload, although some combinations
  of topology and workload will have lower accuracy.
\end{CompactItemize}

\Fig{overview} illustrates the intuition behind its core method, and
\Fig{workflow} depicts its workflow.
The user supplies 1) a description of the topology, as a set of
nodes and links, and 2) the workload, as a set of flows and routes.
In our implementation, we generate the flow list by sampling from the traffic matrix and
the flow size distribution, with inter-arrival times determined by a burstiness
parameter.
%
%
%
Once inputs are supplied, \sys proceeds in several steps:

\Para{Decomposition.} To start, flows are assigned to each link they traverse,
e.g., for a fat tree using ECMP.
%
%
Then, for each link $l$, \sys generates a custom backend simulation with a topology
selected to determine---as accurately as possible---the \emph{contribution of}
$l$ to the end-to-end flow completion times (FCTs) of the flows passing through it.
Each of these backend simulations can run in parallel.

\Para{Clustering.}
Depending on the size of the topology,
there may be tens or hundreds of thousands (or more) of link-level simulations
to perform.
Fortunately, data center topologies exhibit notable symmetries, and industry
has reported that the same is true for many of their
workloads~\cite{fb-network}.
%
\sys can optionally cluster links with similar workloads together.
Only one representative from each cluster need be simulated; the rest of the
link-level simulations are pruned.
Clustering is discussed in more detail in \Sec{clustering}.

\Para{Simulation.} The next step is to simulate all cluster representatives in
parallel.
The decomposition step resulted in a topology and a workload for each
link-level simulation, and we can use any simulation backend.
Our prototype supports two: ns-3 and a custom
high-performance link-level simulator (\Sec{fast-sim}). This allows us
to directly validate our link-level simulator against ns-3. However, other
efficient models, such as fluid flow~\cite{fluid} or machine learned
models could be used here instead, for different tradeoffs of performance
and accuracy.
Each link-level simulation produces a distribution of the delay contributed
by that link to the flow completion time (FCT), bucketed by flow size.
Note this is not the link's
propagation delay---we calculate that contribution directly from the topology.
%
These distributions---described in the next section (\Sec{key-methods})---are
organized according to the original input
topology, as depicted in \Fig{overview}.
Recall that only one representative from each cluster is simulated; every other
link is populated with the distributions of its cluster representative.

\Para{Aggregation.} The last step is to aggregate the link-level results into
estimates for entire paths through the network.
These estimates are also represented as delay distributions.
Conceptually, \sys obtains a delay distribution for a path by convolving
together the appropriate distributions from each of the path's component links.
Since there are multiple distributions per link and potentially many paths
through the network, we do not compute convolutions up-front.
Instead, convolution is done on-demand via Monte Carlo sampling; a by-product
is that we can efficiently produce estimates for individual source-destination pairs,
virtual networks, or classes of service \rev{(\Sec{workload-mixes})}.
To make a single point prediction for a flow taking some path through the network, \sys
uses the flow size to find the appropriate distribution for each link, samples a value from each of them, and
combines them together.  This process is repeated for each flow.
%

\vspace{4pt}
At a bird's-eye view, \sys's method is simple: to accelerate FCT estimates, we
estimate the effect of each link independently and in
parallel.
Then to make predictions about the whole network, we combine the results.
However in our experience, the accuracy of the method hinges tightly on the
quality of the link-level estimates and subsequent aggregation.
For example, when generating the backend simulations, we have observed that
failure to adequately capture pertinent features of the network severely
degrades the quality of \sys's estimates.
Similarly, link-level results must be processed and aggregated with care to
preserve accuracy across all flow sizes.
\Sec{key-methods} describes these techniques in detail.

%% file: methods.tex
\section{Key Methods: Decompose and Aggregate} \label{s:key-methods}

\begin{figure*}[t]
  \centering
  \includegraphics[width=\linewidth]{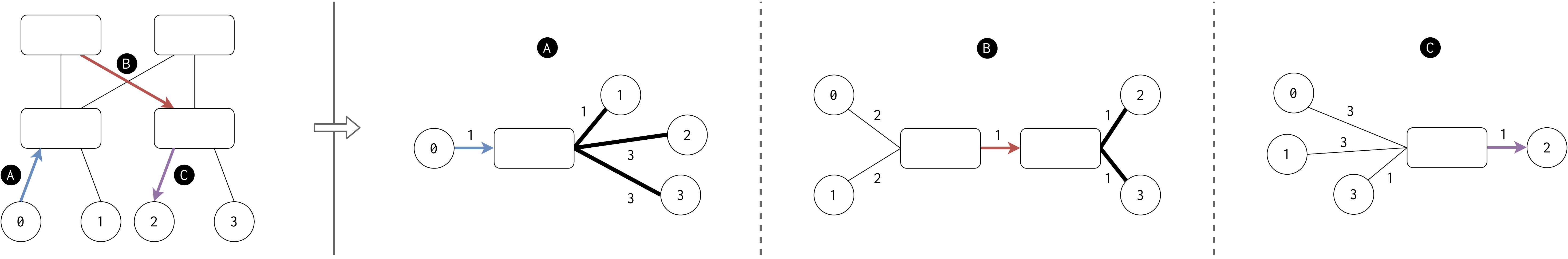}
  \vspace{-6mm}
  \caption{\small
    An illustration of how \sys generates link-level topologies.
    Simulations are unidirectional, and a different topology is used for (A) first-hop links,
    (B) switch-to-switch links, and (C) last-hop links.
    For illustration purposes, each link in the original topology has a propagation delay of one.
    To the left is the original topology; to the right are the corresponding link-level topologies,
    with new propagation delays annotated. Bold lines denote links whose bandwidths have been artificially
    increased during topology generation.
  }
  \label{fig:link-level}
  \vspace{-2mm}
\end{figure*}


Together, the methods for decomposition and aggregation are what enables \sys's
scaling, and while we later engage additional techniques for further
speed-up, they are a byproduct of---and not independent from---these more
essential methods.
Decisions made during this step are also the central determiners of accuracy.
This section describes these processes in detail: how link-level topologies are
generated, how the link-level data are post-processed and stored, and finally
how they are aggregated to produce end-to-end estimates.

\subsection{Generating Link-Level Workloads} \label{s:link-workloads}

To start, \sys associates each link with the flows passing through it.
Since links are bidirectional, there are two sets of flows---and consequently
two link-level simulations---per link.
\sys populates links with flows using flows' routes.
Then for each link and in each direction, the associated flows constitute the
input workload to the link-level simulation.
The sizes and arrival times of the flows pass though unmodified.

\subsection{Generating Link-Level Topologies} \label{s:link-topos}

Once the link-level workloads are in place, we generate the link-level
topologies.
In this step, we think of each link as contributing some amount of delay to
end-to-end FCTs.
Any given flow will accrue these delays at each hop, depending on---for
example---how much bandwidth is available and how much queueing is present.
Highly-loaded links are expected to contribute more delay, while rarely
utilized links will contribute relatively little.

For each link and in each direction, we generate a topology and perform a
simulation using just the flows traversing that link.
Once the simulation is finished, the delay caused by the link for a given flow
is computed by taking the observed
FCT and removing the ideal FCT for that flow size.
(For a flow of size $s$ traversing a link of speed $C$ and propagation delay $l$,
the ideal FCT is $s/C + l$.)
%
This intuitively captures all delays incurred due to queueing,
congestion control, bandwidth sharing, and so on at the target link.

In generating a per-link topology, our goal is to isolate and measure the expected delay
contribution of the target link. A simple but inefficient strategy would
be to use the original
topology, but with only the traffic traversing the target link, without any
cross traffic.  This would be relatively accurate
at measuring the delay contributed by the target link, albeit a bit conservative.
Upstream cross traffic congestion will slightly smooth out downstream
congestion at the target link, and so removing cross traffic would make the
queue distribution at the target link slightly worse than in reality.

Although relatively accurate and parallelizable, simulating every link on the original network
topology would still be inefficient, as packet-level simulation cost is roughly proportional to the
number of packets simulated times the number of hops each packet takes through the network.
Because we run the link simulation separately in each direction on every packet that passes through that link,
this would inflate the aggregate computational cost of the simulation
by a multiplicative factor of roughly half the average network path length---a significant factor for large-scale networks.
Instead, we want to simulate only a small constant number of hops per target link.

An extreme alternative would be to simulate only the target switch queue.
This is inaccurate for two reasons.  First, we need to preserve
end-to-end round trip delays, as these affect the speed of the congestion control
adaptation to congestion or its absence; hosts closer to the target adapt faster than those
farther away.  Second, we need to preserve the
spacing of packets induced by the original topology---a large flow does not immediately dump
all of its data into the queue for the target link; instead, those packets arrive
spaced apart by the edge link capacity. Ignoring this effect would lead to larger
queues and more delay at the simulated link than would occur at that link in the original network.
%

Thus, we construct a topology for each link-level simulation that reflects a performance-accuracy
tradeoff, attempting to capture the most important effects for computing the delay contributed by
the target link.
\Fig{link-level} shows how topologies are minimized.
The generated topology takes one of three shapes, depending on the location and direction
of the target link: (i) a first-hop up-link from a host to a ToR, (ii) a switch-to-switch link in the middle of the network, or (iii) a last-hop downlink from a ToR to a host.

%



Suppose the traffic through the target link
originates from sources $S$ and terminates in destinations $T$.
%
In case \texttt{A} of \Fig{link-level}, we connect the target link directly to each
host in $T$ via a dedicated link.
If the target link is a switch-to-switch link (case \texttt{B}), we remove intermediate hops and connect
the hosts in $S$ directly to the input, and the output directly to the hosts in $T$.
Lastly, if the target link is a last hop (case \texttt{C}), then the
hosts in $S$ are connected directly to the input.
Rewriting the topology in this manner ensures that packets can traverse at most
three hops, regardless of the size of the original topology.

\Para{Modeling round-trip delay.}
Next, we set the link delays in each constructed topology to match the round trip
delays in the original network.
For example, in case \texttt{A} of \Fig{link-level}, the round-trip time between host \texttt{0} and
host \texttt{2} is 8 in both the original topology and the
generated topology, even though \sys has removed intermediate hops between the
switch and host \texttt{2}. \Fig{link-level} is meant as illustrative;
as with ns-3, \sys can model arbitrary round-trip delays.

In data center networks, congestion controllers play a large role in
determining the extent to which longer flows yield throughput to benefit the
latency of short flows.
Most algorithms such as DCTCP~\cite{dctcp}, DCQCN~\cite{dcqcn}, and
TIMELY~\cite{timely} are \emph{end-to-end} in the sense that sources adjust
their sending rates based on feedback echoed from destinations~\cite{bfc}.
With an end-to-end control loop, a source
must wait an entire round-trip time (RTT) before being able to adapt its
sending rate based on congestion feedback, resulting in longer queue lengths
with higher RTTs. Thus, correctly modeling RTTs is essential to correctly modeling
queue dynamics.

\Para{Selecting link bandwidths.}
In some cases, we
artificially increase the bandwidth of downstream links to ensure that
they do not artificially add congestion. We say such links are \emph{inflated}.
For example, in cases \texttt{A} and \texttt{B} of \Fig{link-level}, the
bandwidths of the last-hop links are inflated.
We want any queueing to be due to the target link and not the downstream link.
By inflating downstream links, we remove store and forward delay
(a small packet following a large packet
would otherwise need to queue for the downstream link); it also addresses the
case where core links are fatter than downstream links.
Queueing at the downstream link itself is accounted for in case \texttt{C}.
By contrast, we do {\em not} inflate first-hop links in cases
\texttt{B} and \texttt{C}, as this would enable a long flow to arrive at
the target link at a higher rate than it would in practice.

A cluster of sources sending simultaneously through an oversubscribed
top-of-rack (ToR) switch in the original network will be throttled beyond
what is implied by the edge link capacity.
To improve simulation speed, we ignore this effect and are therefore
slightly conservative in our estimates for oversubscribed networks.

\Para{Correcting for ACK traffic.} Since \sys only simulates one direction at a
time, we must account for the load induced by acknowledgments due to traffic
in the reverse direction.  This is usually small, but can be significant at high load
and where average packet size is small.  Instead of modeling ACK traffic in detail,
we apply a simple rule, mechanically reducing the forward bandwidth on each simulated
link by the average volume
consumed by ACKs for flows in the opposite direction over the course of the simulation.
This correction is applied to all links but is most necessary for the target
link.  Note that \sys does not account for extra delay caused by ACK jitter on the reverse path;
this could be an issue when applying our ideas to networks with bandwidth asymmetry between forward and reverse paths~\cite{reversepath}.

\subsection{Post-Processing Link-Level Results} \label{s:post-processing}

Each link-level simulation produces an FCT for each flow in the link-level
workload, and these FCTs are used to compute delays.
Recall from \Sec{link-topos} that the delay is just the observed FCT minus the
ideal FCT on an unloaded network.  For each flow, we could, theoretically, estimate the end-to-end delay
as some function of the delay contributed by each link for that flow. We discuss how that function works
in \sys, along with its sources of bias, later in this section.

First, we address a different issue. Recall that we cluster similar links together (\Sec{clustering})
so that we only simulate the flows through a single representative link for each cluster of links.
Thus, to compute the end-to-end delay for a particular flow, we take a sample
from the delay distributions at each hop in the path, or from the hop's standin representative.

In post-processing the link-level results and constructing these distributions, our
primary objective is to support accurate estimates for \emph{all flow sizes.}
It is not enough to produce the correct FCT distribution across the
entire workload; we must also accurately estimate the FCT distribution for
short flows containing just a few packets as well as for long flows that last for
hundreds of round trips.
This extra requirement necessitates some post-processing before distributions
can be constructed.
Here we describe how this is done.

\Para{Packet-normalized delay.} Maintaining accuracy across all flow sizes
would not be possible if we used delays directly.
For example, long flows, which may experience variations in their bandwidth
share over time, will almost always experience more absolute
delay than short flows.

As a start, we can address this by normalizing delays by flow size: after
computing the delay for a particular flow, we can then divide the delay by the
flow's size in packets.
We call the resulting metric the \emph{packet-normalized delay}, and it has the
intuitive interpretation of summarizing the flow's average delay per packet.
Link-level distributions are constructed from packet-normalized delays rather
than absolute delays.
We normalize by the number of packets instead of the number of bytes because
flows are discretized into---and therefore delays are incurred by---packets.
Further, normalizing by the number of bytes loses accuracy for small flows,
especially those smaller than the maximum packet size.
\revv{
  Both 50 and 100 byte packets are equally delayed in the tail if they arrive
  in the switch queue behind a jumbo (9 KB) packet~\cite{jumboframes}.
}{
  For example, a 10 byte packet would be delayed by the same amount as would
  a 100 byte packet if it arrived in the switch queue just behind a jumbo (9
  KB) frame~\cite{jumboframes}.
}

\Para{Bucketing distributions.} Even with packet-normalized delays, we should
still expect long flows to have different delay distributions than short flows.
The FCT of a long flow is mainly determined by the throughput it achieves,
while the FCT of a short flow depends on how much queueing it encounters.
Further, congestion control algorithms trade the throughput of long flows for
the latency of shorter ones to varying degree.
An aggressive congestion control algorithm could try to keep queues
near-empty~\cite{hpcc}, resulting in smaller short-flow delay and larger
long-flow delay.

To ensure that estimates for different flow sizes are accurate, it is necessary
to sample each packet-normalized delay from the appropriate distribution.
%
We bucket the distribution of packet-normalized delays by flow
size. Buckets need to contain enough samples to form statistically meaningful distributions, but
they should also be small enough so that the values come from
flows with similar delay characteristics (\ie similarly-sized flows).

\sys uses a simple bucketing algorithm.
In brief, we start with a packet-normalized delay per flow, and we sort them
according to flow size.
Then, starting with the shortest flow, we begin populating buckets.
For each bucket $b$, let $\text{maxf}_{b}$ and $\text{minf}_{b}$ be the maximum
and minimum flow sizes associated with $b$, respectively, and let $n_{b}$ be
the number of elements in $b$.
Each bucket $b$ apart from the last one is locally subject to two constraints
\vspace{-1mm}
\begin{equation*}
  n_{b} \geq B
  \qquad \text{and} \qquad
  \text{maxf}_{b} \geq x * \text{minf}_{b},
  \vspace{-1mm}
\end{equation*}
%
for some choice of $B$ and $x$.
Globally, \sys also ensures buckets are contiguous and non-overlapping.
For any bucket, once the two local constraints are satisfied, \sys begins
populating the next bucket, and the final bucket is assigned whatever elements
remain.

In practice, we find $B = 100$ and $x = 2$ works well.
Data center workloads have heavy-tailed flow size distributions in which short
flows arrive much more frequently than long ones.
With these parameters, the first buckets will have size boundaries that are
approximately powers of two, and as flows get larger, buckets will cover larger
and larger ranges.
This is the desired behavior.
Intuitively, a \rev{queueing-sensitive} 1 KB flow should not be grouped
with a \rev{throughput-sensitive} 1 \revv{MB}{GB} flow, but a 1 GB flow
can be \revv{safely}{} grouped with a 10 GB flow \revv{because we expect them to have similar
  delay contributors}{provided the distribution of throughput is stable on long timescales}.
\rev{
  Accuracy across different flow sizes at finer or coarser resolution can be
  achieved by modulating $x$. We examined sensitivity to the number of buckets by decreasing
   $x$ for selected experiments and found no meaningful change in the predicted distributions.
}

\vspace{4pt}

In summary, each link-level simulation produces FCTs, and these FCTs are used
to construct bucketed distributions of packet-normalized delay.
Since different links have different workloads, bucketing is performed on a
per-link basis.
This means that the links in any given path are likely to have different bucket
sizes with different flow size ranges.
In the next subsection (\Sec{aggregation}) we describe how the data are
aggregated.

\subsection{Aggregating Link-Level Estimates} \label{s:aggregation}

\begin{figure}[t]
  \centering
  \includegraphics[width=0.85\linewidth]{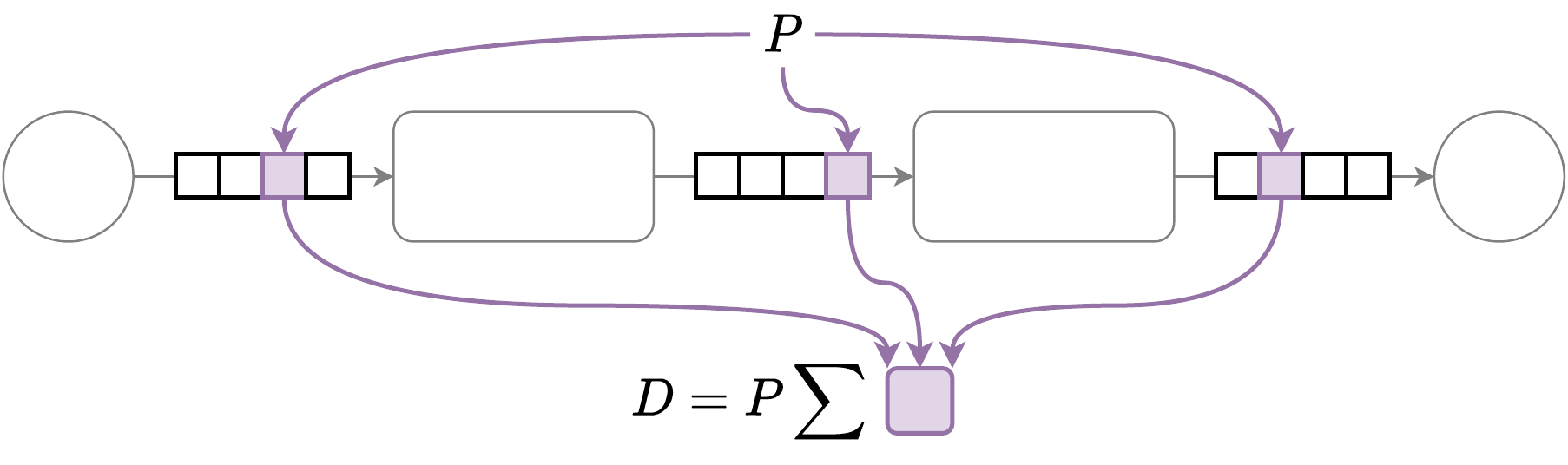}
  \vspace{-4mm}
  \caption{\small
    An illustration of how \sys aggregates link-level results into a
    path-level point estimate. \sys samples a packet-normalized
    delay (\Sec{post-processing}) from each link along the path, and
    combines these to estimate the end-to-end absolute delay $D$.
  }
  \vspace{-4mm}
  \label{fig:aggregation}
\end{figure}

For any given range of flow sizes, the final distribution of
(packet-normalized) delay for any path through the network can be estimated by
selecting an appropriate distribution from each component link and then
performing an n-ary convolution.
However, the efficiency of this step must be considered.
Since there are multiple distributions per link and potentially many paths
through the network, performing all convolutions up front and storing one
path-level distribution per path, per flow-size range would be costly in space
and in time.

To avoid these costs, \sys uses an on-demand sampling strategy to perform the
convolution.
Recall that the simulation step resulted in bucketed distributions of
packet-normalized delay per link, organized in a graph isomorphic to the
original topology.
\sys makes this graph a queryable object that is capable of supporting point
estimates.
Given a size, a source, and a destination, \sys computes a path from the source
to the destination and uses the size to select a distribution per-link.
Then, one packet-normalized delay is sampled from each distribution and the
results are subsequently combined into a point estimate.
Suppose there are $n$ hops and let $D^*_1, D^*_2, \ldots, D^*_n$ be the sampled
packet-normalized delays.
Then, the end-to-end absolute delay $D$ is computed as
\vspace{-1mm}
\begin{equation*}
  P \sum_{i=1}^{n} D^{*}_{i}
  \enspace = \enspace \sum_{i=1}^{n} D^{*}_{i}P
  \enspace = \enspace \sum_{i=1}^{n} D_{i}
  \enspace = \enspace D,
  \vspace{-1mm}
\end{equation*}
%
where $P$ is the input flow size in packets and $D_i$ is the absolute delay for
hop $i$.
\Fig{aggregation} illustrates this process.
Finally, to obtain a distribution of end-to-end delay estimates, we need only
sample enough point estimates for the desired flow size range and source
destination pairs.

\rev{
  \subsection{Primary Source of Speedup} \label{s:speedup-source}
}

\sys speeds up large network simulations by considering the effect of each link
in isolation, allowing it to scale in the size of the simulated network and the
number of processing cores.
Although the link is the unit of decomposition, \sys's scaling ability is
determined not by the total the number of links, but rather by the
\emph{fraction of total packets traversing any link.}
In other words, \sys's speed-up depends on the number of busy links and how
well the load is balanced among them.
This explains why \sys is most suited for large data center networks, where the
total workload comprises many source destination pairs with many paths between
them.
If a network traffic is
heavily skewed such that most of the workload traverses only a few paths,
the amount of speedup will be limited.

\subsection{Primary Sources of Error} \label{s:error-sources}

To balance accuracy and performance, \sys makes a number
of approximations, with some having more of an effect on accuracy than
others.
Here we catalog some of the main sources of error, describing 1) how we expect
the errors to manifest and 2) what modifications, if any, could be made to address them.

\Para{Bottleneck fan-in.} To simulate a given target link in the network, \sys constructs a topology
that connects all of the source nodes feeding traffic directly into that target.
In practice, of course, there would be multiple stages of fan-in, and that fan-in would tend to
spread out any burst of arriving flows due to upstream bandwidth capacity constraints.
Any target
link would experience slightly less queueing and less congestion in reality than in \sys.
Of course, \sys also simulates the upstream link; because it is closer to the sources,
its traffic and queueing behavior would be a closer model to what would happen in a
full network-wide simulation.

Because \sys sums the delay contributed by each hop along a flow's path, the lack of
fan-in will tend to slightly overestimate the delays caused by downstream links.
Put another way, any delay induced by fan-in constraints is counted twice---once
when we simulate the upstream link and again when we simulate the downstream link.
In our evaluation, accuracy
is slightly lower for networks with higher degrees of oversubscription, as we would expect.
We could potentially remove this inaccuracy by including the upstream fan-in as part of the
topology for each link simulation. Since simulation time is proportional to the number of hops,
this would decrease individual link simulation efficiency by a small but significant factor.

\Para{Lack of traffic smoothing.}
Similarly, any cross-traffic that shares a portion
of a path with traffic destined for the target link will tend to smooth out traffic
before it reaches the target.  \sys does not include any cross-traffic in its per-link
simulation, making it slightly overestimate the queueing delay at the target link.
Assuming the simulation is stable---that the arrival rate does not exceed the
service rate for any link---the target link will experience the correct
long-term average rate, but without as much smoothing as would happen in practice.
We see evidence of this effect in our evaluation, where error is slightly larger
for workloads with a predominance of short flows which would benefit more from smoothing.
Of course, correctly modeling the effect of cross-traffic on the traffic arriving
at a downstream link would be difficult to accomplish without reverting to a full network simulation.

\Para{Link-level independence.} A more fundamental approximation is that
link-level simulations are treated independently.
This technique enables wholesale parallelization, but its accuracy depends on
the amount of correlation between the traffic intensities on the various
hops along the path.
The more correlated the traffic, the more error \sys's method produces.

\revv{
  First, consider the effect \sys's method will have on the FCT of a long flow.
  If the intensity of the coresident traffic at each hop of the flow's path is
  independent, then the flow's packets are unlikely to be backlogged on multiple
  links at the same time---queueing, and the resulting end-to-end congestion
  response, will happen at one link at a time.
  Therefore, we would expect the flow to sometimes be backlogged on one link, and
  at other times backlogged on another.
  In this case, summing over the delay contributions of each link---as \sys
  does when convolving the distributions---results in a good estimate
  of end-to-end delay.
  On the other hand, if the intensity of the coresident traffic were perfectly
  positively correlated, the flow's packets could be backlogged on multiple links
  during the \emph{same} time interval.
  As before, the aggregation method would add the delays contributed by each
  link; however in this case, doing so would produce an overestimate, because
  during that time interval, only one bottleneck would contribute to end-to-end
  delay.

  For short flows, a primary effect of correlated traffic is to alter the
  probability that a flow will encounter queueing.
  Consider a single-packet flow that traverses two hops, both with load $l$.
  If the traffic intensity over time along the two hops is independent, the
  probability that the flow will encounter \emph{no queueing} is simply
  $(1-l)^2$.
  However, if both hops tend to have queueing at the same time, then the
  probability of encountering no queueing is closer to $1-l$.
}{
  Since \sys produces estimates by convolving delay distributions
  (adding independent random variables), full
  accuracy requires the mutual independence of delays among the links in
  every path.
  Consider a single-packet flow that traverses two hops, both with load $l$.
  If the delays along the two hops are independent, the probability that the
  flow will encounter \emph{no queueing} is simply $(1-l)^2$.
  However, if both hops tend to have queueing at the same time (\ie if the
  traffic intensities and therefore the delays are correlated), then that
  probability is closer to $1-l$. 
  %
  Since \sys does not distinguish between these two scenarios, the
  difference is not reflected in its estimates.
}

In very large networks with thousands of hosts and paths, and with realistic
workloads, we expect the effects of correlation to be small.
A basic result of queueing theory is that under some circumstances it is
possible to analyze queues independently, even when the output of one queue
connects to the input of another, so that queue behaviors are obviously
correlated. One view of our work is that we are empirically observing that data
center networks approximately admit product-form solutions for their
equilibrium state queue distributions under realistic workloads.

However, some networks
use PFC~\cite{dcqcn} to reduce packet loss due to go-back N error handling in some RDMA network
interface cards. Because PFC suffers
from head-of-line blocking, PFC can cause correlated congestion across multiple links, and so \sys would not
be a good choice for modeling such networks. If correlation is a problem, we could potentially measure
the degree of correlation and apply a correcting factor during the convolution step, but we leave that
for future work.

\Para{One bottleneck at a time.}
\revv{
  Finally, in the convolution function,
  we sum delays contributed by each link in the path rather than use a more
  complicated function. The delay for a flow of a few packets is
  well-modeled as the sum of the delay that those packets experience at each hop.
  However, summing delays is less good for long flows that encounter simultaneous
  cross-traffic congestion at multiple points along the path.  In this case, \sys
  will ``double count'' those delays---the congestion control
  protocol will (generally) back off based on the most congested link rather
  than based on the sum of the simultaneous congestion.
}{
  Estimating the performance of long flows comes with an additional
  difficulty which is also exacerbated by correlated delays.
  While a single packet flow can only reside in one queue at a time, a long
  flow can be backlogged on multiple links \emph{at the same time.}
  Depending on the specific congestion control mechanism, the throttling
  back of a long flow (the delay it experiences) is typically \emph{not}
  the sum of the delays it would experience on individual links (as \sys approximates), but
  rather only the delay caused by the true (instantaneous) bottleneck. 
  Since \sys sums all delays, it will overestimate the end-to-end
  delay for the long flow that encounters simultaneous cross-traffic congestion at
  multiple points along its path. 
  In summary,
}
\sys is more accurate when the congestion is episodic and temporary, appearing
at different links at different times, \rev{and less accurate when congestion is persistent
across multiple edge and core links of a given path.}

Congestion on any link (and therefore simultaneous congestion on multiple links) becomes
more common with higher network load, and we see this effect in our evaluation.
We can potentially correct for this bias by using a more complex function
for combining link delays when overall network utilization is high.
Because network operators are often willing to over-provision their network hardware
to reduce application tail latency, this is rare in practice.  For example, some recent
end-to-end congestion protocols, such as Homa~\cite{homa}, simply assume that network congestion
predominantly occurs at the last hop of each path.  We do not make such an assumption;
we handle congestion equally wherever it might occur.  However, we do assume that
congestion events are not persistent and network wide.

\vspace{4pt}

\revv{Each of these approximations is}{Our approximations are} biased toward
producing overestimates rather than underestimates, because we expect network operators
to be more sensitive to over-promising tail behavior, even if that
comes at the cost of being too conservative with respect to capacity planning.
Additional analyses on the errors induced by these approximations can be found
in the appendix (\Sec{app-error-sources}).

\section{Complementary Methods} \label{s:other-methods}
The previous section described how we decompose a single large network simulation into many small,
independent ones that can be executed in parallel and later combined. This section describes
additional optimizations that reduce, cluster, and prune these link-level simulations
for better computational efficiency. These reduce the number of cores needed
to simulate a given network within some time bound, or equivalently, the execution time
on a single server machine.

\subsection{Fast Link-Level Simulation} \label{s:fast-sim}
By far the largest computational cost in \sys are the link-level simulations.
Initially we used ns-3 as our link-level backend.  However,
as a general-purpose simulator, ns-3 is designed to support arbitrary protocols with arbitrary
extensions, all the way down to hardware models. This is more flexible but
means that every packet in ns-3 generates events at every host, queue, and
link---as well as throughout the hosts' modeled network stacks.

Instead, we implemented a custom and minimal simulator optimized for high fidelity
single link simulation.
This backend only models the workload, topology, queueing, and
congestion control.
For congestion control, our prototype implements DCTCP's core
algorithm~\cite{dctcp} in a few tens of lines of code.  For example,
we do not need to model the mechanism for carrying ECN bits from switches
back to endpoints. Switching to a custom simulator speeds up the individual
link simulations by roughly an order of magnitude, with negligible loss of accuracy.
Reducing the simulation time of the worst case (most congested) link also reduces the
critical path dramatically. If more simulation features
are needed, \sys can use ns-3 at the cost of using more cores.
%

\subsection{Clustering and Pruning Simulations} \label{s:clustering}

\begin{algorithm}[t]
  \small
  \caption{Greedy link clustering \label{alg:greedy}}
  \begin{algorithmic}[1]
    \Let{unclustered}{\Call{AllLinks}{}} \Comment{links here are unidirectional}
    \Let{clusters}{[]} \Comment{list of list of links}
    \While{\textbf{not} {\Call{Empty}{unclustered}}}
    \Let{members}{[]} \Comment{new cluster}
    \Let{representative}{\Call{PopFirst}{unclustered}}
    \State \Call{Push}{members, representative} \Comment{with initial member}
    \For{candidate \textbf{in} unclustered} \Comment{find other members}
    \Let{rfeature}{\Call{Feature}{representative}}
    \Let{cfeature}{\Call{Feature}{candidate}}
    \If{\Call{IsCloseEnough}{rfeature, cfeature}}
    \State \Call{Push}{members, candidate} \Comment{new member}
    \State \Call{Remove}{unclustered, candidate}
    \EndIf
    \EndFor
    \State \Call{Push}{clusters, members}
    \EndWhile
    \State \textbf{return} clusters
  \end{algorithmic}
\end{algorithm}

Lastly, we recall that \sys's decomposition results in two simulations per
link: one in each direction (\Sec{link-workloads}).
On a large-scale 6,144-host topology we use for evaluation, there are over
9,000 links, and therefore over 18,000 simulations generated.
Fortunately, data center topologies commonly induce symmetries that render some
of these simulations redundant.
For example, up-links in the same ECMP grouping can be assumed to have the same
characteristics and traffic patterns.
Furthermore, the workloads themselves may also induce symmetries due to
communication patterns and load balancing~\cite{fb-network}.

We can take advantage of these symmetries by clustering links that carry similar
traffic and only simulating one representative from each cluster.
Then, in each cluster, all links inherit the delay distribution produced by the representative link.
\sys's clustering requirement is quite specific, which limits the range of
popular clustering algorithms that can be used.
Let $l_1, l_2 \in L$ be any two link-level simulations, and let $d : L \times L
  \rightarrow \mathbb{R}$ be a distance function.
Ideally,
\vspace{-1mm}
\begin{equation*}
  l_1 \text{ and } l_2 \text{ are clustered together} \iff d(l_1, l_2) < \epsilon,
  \vspace{-1mm}
\end{equation*}
where $\epsilon$ is some bound.
The left-to-right direction preserves accuracy; the
right-to-left supports efficiency.
Most centroid-based and density-based clustering algorithms aren't designed to
provide the left-to-right property. Instead, \sys uses \Alg{greedy}.
This algorithm greedily clusters simulations together, using a distance function that
predicts which links will have similar delay profiles. In our prototype, we
check that the link flow size and inter-arrival time distributions---as well as
their load levels---are close. We find this provides a reasonable tradeoff
between efficiency and accuracy, but users can turn off the optimization at the cost
of using more cores. Further details about the clustering can be found in the appendix (\Sec{more-clustering}).
%
%
%


%% file: evaluation.tex
\section{Evaluation} \label{s:eval}

\sys's goal is to quickly estimate tail latencies for a variety of large data
center networks and workloads.
In evaluating \sys, we would like to assess 1) \sys's accuracy and performance
at the scale of thousands of hosts, and 2) how accuracy is affected by a wide
range of variables over the workload and the topology.

Our strategy is as follows.
Using workloads extracted from industry datasets, we start with a
384-rack, 6144-host topology to evaluate \sys's speed and accuracy in one
scenario at scale.
Then, to evaluate nearly 200 other topology and workload scenarios, we
downsample the workload so that it can run on a smaller 256-host topology.
This allows us to run enough ns-3 simulations quickly enough to perform a detailed sensitivity
analysis.

\rev{To more clearly illustrate sources of error in \sys, we also construct and evaluate \sys on synthetic workloads 
on a small-scale parking lot topology in Appendix \Sec{app-error-sources}.}
%

\subsection{General Setup} \label{s:eval-setup}

\begin{figure*}[t]
    \centering
    \begin{subfigure}[t]{0.342\linewidth}
        \centering
        \includegraphics[width=\textwidth]{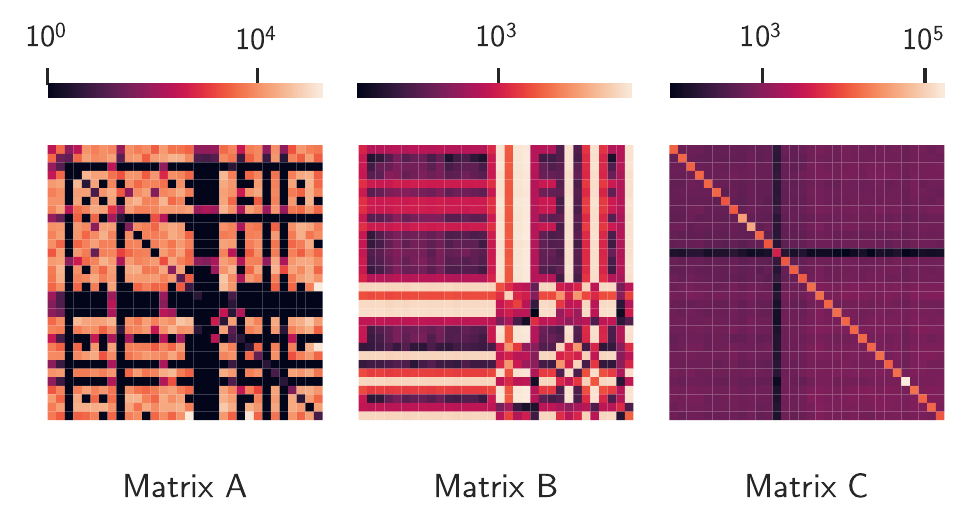}
        \vspace{-6mm}
        \caption{\small
            Traffic matrices (32-rack sample)
        }
        \label{fig:matrices}
    \end{subfigure}
    \hfill
    \begin{subfigure}[t]{0.241\linewidth}
        \centering
        \includegraphics[width=\textwidth]{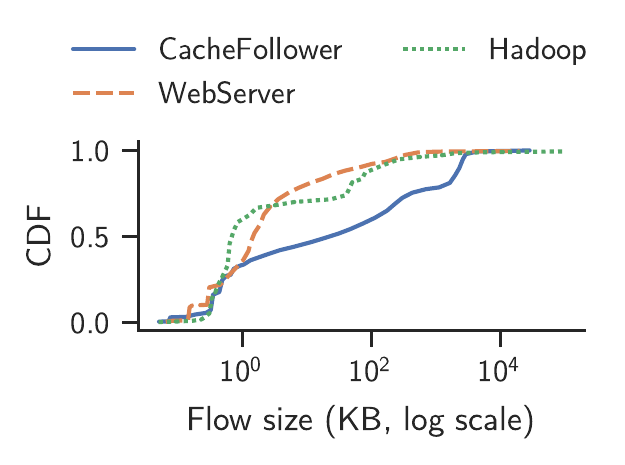}
        \vspace{-6mm}
        \caption{\small
            Flow size distributions
        }
        \label{fig:flow-sizes}
    \end{subfigure}
    \hfill
    \begin{subfigure}[t]{0.40\linewidth}
        \centering
        \includegraphics[width=\textwidth]{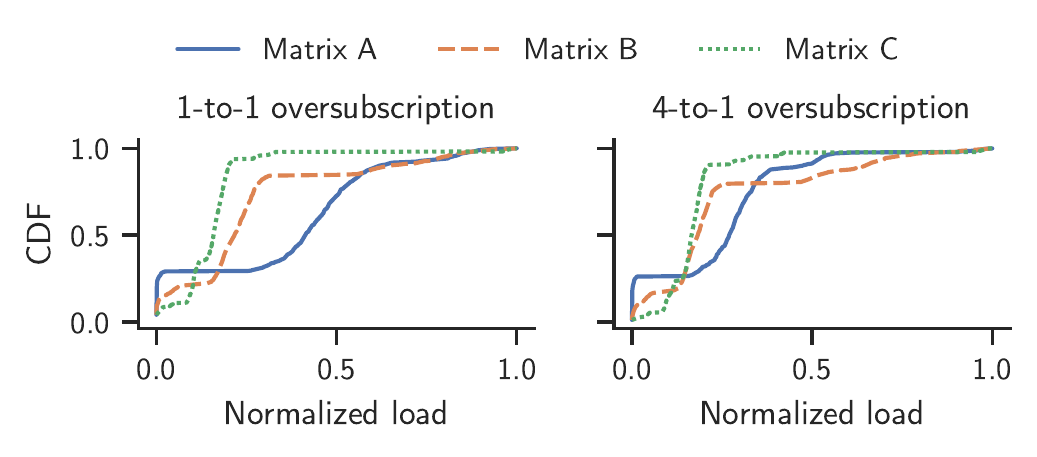}
        \vspace{-6mm}
        \caption{\small
            Normalized link load distributions
        }
        \label{fig:link-loads}
    \end{subfigure}
    \vspace{-3mm}
    \caption{\small
        In the evaluation, we model workloads using data from Roy \etal's study
        of Meta's data center network~\cite{fb-network}.
        The traffic matrices in \Fig{matrices} are extracted from the
        accompanying dataset, and the flow size distributions in
        \Fig{flow-sizes} are estimated from the published data.
        Lastly, for a given topology, the distribution of link loads depends on
        1) the traffic matrix and 2) the degree of oversubscription.
        \Fig{link-loads} shows the link loads induced by the matrices in
        \Fig{matrices} on two 32-rack topologies with different
        overprovisioning.
        The x-axis is normalized to the maximum link load.
    }
    \vspace{-3mm}
    \label{fig:workload}
\end{figure*}

\begin{table}
    \small
    \centering
    \begin{tabular}{lcc}
        \toprule
        Variant   & Clustering? & Link-level backend \\
        \midrule
        \sys      & No          & custom             \\
        \sys/C    & Yes         & custom             \\
        \sys/ns-3 & No          & ns-3               \\
        \sys/inf  & ---         & custom             \\
        \bottomrule
    \end{tabular}
    \caption{\small
        The \sys variants under consideration.
        \sys/inf is a variant that assumes infinite cores and memory.
    }
    \vspace{-10mm}
    \label{tab:variants}
\end{table}

Each scenario we consider has six components: 1) a topology size, 2) an
oversubscription factor, 3) a traffic matrix, 4) a flow size distribution, 5) a
burstiness level, and 6) a maximum load level.
Here, we briefly describe how these are specified and configured.
We also discuss which \sys variants we will assess and how we establish a
baseline.

\Para{Topology and oversubscription.} To mimic an industry topology, our
topologies are modeled after Meta's data center
fabric~\cite{fb-fabric}.
In brief, there are three layers of switches: hosts connected to a top-of-rack
switch (ToR) with 10 Gbps links constitute a \emph{rack}, racks connected to
each other via fabric switches with 40 Gbps links constitute a \emph{pod}, and
pods connected to each other via spine switches with 40 Gbps links constitute a
\emph{cluster}.
Spine switches are organized in \emph{planes}.
%
We can modulate the size of a topology (corresponding to a cluster)
by adjusting the number of pods, the number of racks per pod, and the number of
hosts per rack, and we can modulate the oversubscription factor by adjusting
the number of spines per plane.

\Para{Traffic matrices.} The traffic matrices are extracted from the datasets
accompanying Roy \etal's study of Meta's data center network~\cite{fb-network}.
The data only allow us to construct reliable rack-to-rack matrices.
When sampling workloads, we use the matrices to generate rack-to-rack traffic,
but once a rack is chosen, we select its hosts uniformly at random.
This may bear semblance to reality: according to Roy \etal, Meta's racks
typically only contain servers in the same role, and load balancing is used
pervasively.
We use traffic matrices from three different clusters: a database cluster
(matrix A), a web server cluster (matrix B), and a Hadoop cluster (matrix C).
\Fig{matrices} shows 32-rack samples of the matrices.

\Para{Flow sizes and burstiness.} We use three flow size distributions,
estimated from published data in Roy \etal's study~\cite{fb-network}.
These are reproduced in \Fig{flow-sizes}.
For inter-arrival times, we use the log-normal distribution to model bursty
traffic, and we modulate the burstiness by adjusting the log-normal shape
parameter $\sigma$.
For low burstiness, we select $\sigma = 1$, and for high burstiness, we choose
$\sigma = 2$.

\Para{Maximum load level.} When setting a load level, we ensure that the offered rate is less than the link
capacity for each link by specifying the maximum load level that any link can have.
Note that a given maximum load level may result in
different link load distributions, depending on the traffic matrix and the
topology.
\Fig{link-loads} shows the distribution of normalized link loads on a 32-rack
topology with the traffic matrices in \Fig{matrices} and two different
oversubscription factors.
When describing how loaded a topology is, we will usually specify
the average load of the top 10\% most loaded links.

\Para{\sys variants and baseline.} To establish a baseline for \sys's accuracy
and performance, we use ns-3 with the optimized build profile.
%
%
We also consider several \sys variants, summarized in \Tab{variants}.
By default, \sys uses the custom link-level backend (\Sec{fast-sim}) with
clustering turned off.
%
%
This expresses a lower bound on \sys's expected speed-up given a particular
machine.
\sys/C adds clustering to the default variant using the methods described at
the end of \Sec{clustering}, and \sys/ns-3 replaces the default's custom
backend with ns-3.
Lastly, \sys/inf provides an estimate of \sys's performance given infinite
cores and infinite memory, computed by adding the run time of the longest
link-level simulation to the fixed costs of network setup and convolution
sampling.
This represents an upper bound on the \sys's achievable performance.
All performance measurements are taken on a 32-core AMD Ryzen Threadripper
3970X.

\subsection{Analysis on a Large-Scale Network} \label{s:large-scale}

\begin{figure*}[t]
    \centering
    \includegraphics[width=\linewidth]{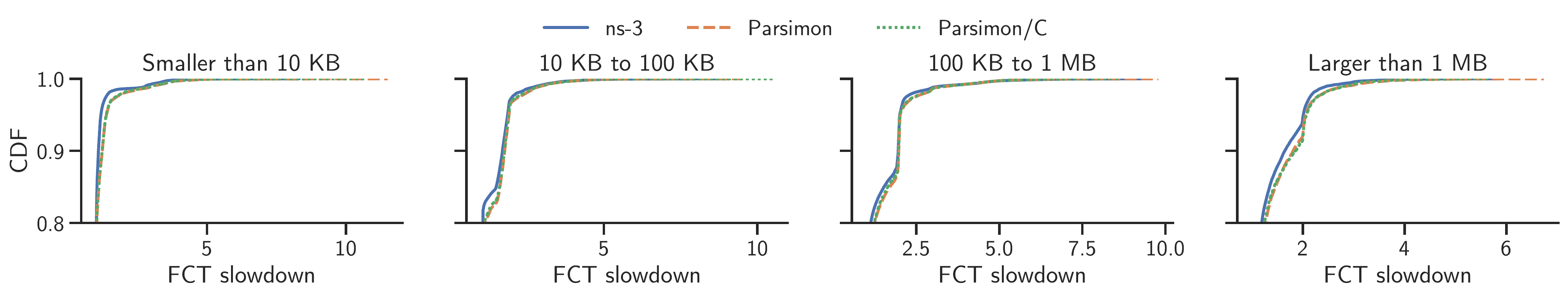}
    \vspace{-7mm}
    \caption{\small
        CDFs of FCT slowdown estimated by ns-3 and two \sys variants (note the
        y-axis).
        On a large network with 6,144 hosts, an industry traffic matrix (matrix
        B), and 2-to-1 oversubscription in the core, \sys's latency estimates
        are similar to those produced by full-fidelity simulation.
        \Tab{large-scale} shows the performance of each estimator.
    }
    \label{fig:large-scale}
    \vspace{-2mm}
\end{figure*}

\begin{table}
    \small
    \centering
    \begin{tabular}{p{2cm}>{\raggedleft\arraybackslash}p{1.6cm}>{\raggedleft\arraybackslash}p{1.6cm}}
        \toprule
        Estimator & Time        & Speed-up     \\
        \midrule
        ns-3      & 10h 48m 26s & ---          \\
        \sys      & 4m 13s      & $154\times$  \\
        \sys/C    & 1m 19s      & $492\times$  \\
        \sys/inf  & 21s         & $1864\times$ \\
        \bottomrule
    \end{tabular}
    \caption{\small
        Running times and speed-up of \sys variants for five seconds of
        simulated time on a large oversubscribed network with thousands of
        hosts.
        We find that \sys estimates latencies orders of magnitude faster than
        does ns-3.
        If there is ample opportunity for clustering or if there are infinite
        compute resources, speed-up is substantially further increased.
        Measurements were taken on a 32-core machine.
    }
    \label{tab:large-scale}
    \vspace{-8mm}
\end{table}

Here we evaluate \sys's accuracy and performance on a 384-rack, 6144-host
topology.
The topology has eight pods, 48 racks per pod, and 16 hosts per rack, with
2-to-1 oversubscription.
For the workload, we use matrix B, the WebServer flow size distribution, and
high burstiness ($\sigma=2$).
We set a maximum link load of about 50\%, which gives the 100 most loaded links
an average load of 32\%, and the top 10\% most loaded links an average load of
about 15\%.
We configure all simulations to run for five seconds of simulated time.
To establish a baseline, we first run the scenario in ns-3, then we run the
scenario in \sys and \sys/C (see \Tab{variants}).
Due to memory constraints we omit \sys/ns-3 here, but we
include its analysis at smaller scale in \Sec{sensitivity}.

\Fig{large-scale} shows the accuracy of \sys relative to ns-3 across four flow
size bins.
We find that across all bins, both variants accurately estimate tail latencies.
If we consider all flow sizes together, we find that \sys and \sys/C overestimate
the p99 FCT slowdown by 8.8\% and 7.5\%, respectively.

\Tab{large-scale} shows the running time and speed-up for each
\revv{
    estimator.
}{
    estimator, which includes topology generation and convolution sampling
    overheads where applicable.
}
While ns-3 took nearly 11 hours, \sys without clustering took four
minutes and 13 seconds, for a speed-up of $154\times$.
If we turn clustering on by using \sys/C, the running time is further reduced
to one minute and 19 seconds, for a speed-up of $492\times$.
\footnote {
    We advise caution both in interpreting this number and in generalizing it
    to scenarios at large.
    While our workloads are modeled after industry data, they are still
    synthetic.
    There may be more or less opportunity to cluster and prune link-level
    simulations, depending on the structure of real workloads and the quality
    of the clustering algorithm.
}
In this case, only 25\% of links were simulated; the rest were pruned.
Lastly, \sys/inf estimates \sys's best possible performance given infinite
compute resources.
The longest-running single-link simulation took 11 seconds, and with the
additional 10 seconds required for network setup and convolution sampling, the
fastest projected running time is 21 seconds.

We chose an oversubscribed topology to slightly disadvantage \sys's method, as
oversubscription can lower \sys's accuracy.
%
%
\Sec{sensitivity} analyzes the effect of oversubscription in more detail.
We also ran the above experiment on a topology without oversubscription,
which for the same maximum load setting increased the top 10\% average link
load from 15\% to 25\%.
We found \sys's p99 accuracy improved from 9\% to about 7\%, while \sys/C's
accuracy remained approximately the same.
However, because aggregate load increased, ns-3 took 27 hours for five seconds
of simulated time, and speed-ups for \sys, \sys/C and \sys/inf were
$152\times$, $872\times$, and $3487\times$, respectively.
\sys/C benefited from the increased number of links in each ECMP grouping,
allowing it to prune 85\% of the link-level simulations.

\subsection{Sensitivity Analysis at Small Scale} \label{s:sensitivity}

\begin{table}
    \small
    \centering
    \begin{tabular}{p{3.2cm}l}
        \toprule
        Parameter              & Sample space                            \\
        \midrule
        Oversubscription       & 1-to-1, 2-to-1, 4-to-1                  \\
        Traffic matrix         & Matrix A, Matrix B, Matrix C            \\
        Flow size distribution & CacheFollower, WebServer, Hadoop        \\
        Burstiness             & Low ($\sigma = 1$), High ($\sigma = 2$) \\
        Max load               & 26\% to 83\% (continuous range)         \\
        \bottomrule
    \end{tabular}
    \caption{\small
        The sample space for the sensitivity analysis in \Sec{sensitivity}.
    }
    \label{tab:sample-space}
    \vspace{-3mm}
\end{table}

\begin{figure}[t]
    \centering
    \includegraphics[width=\linewidth]{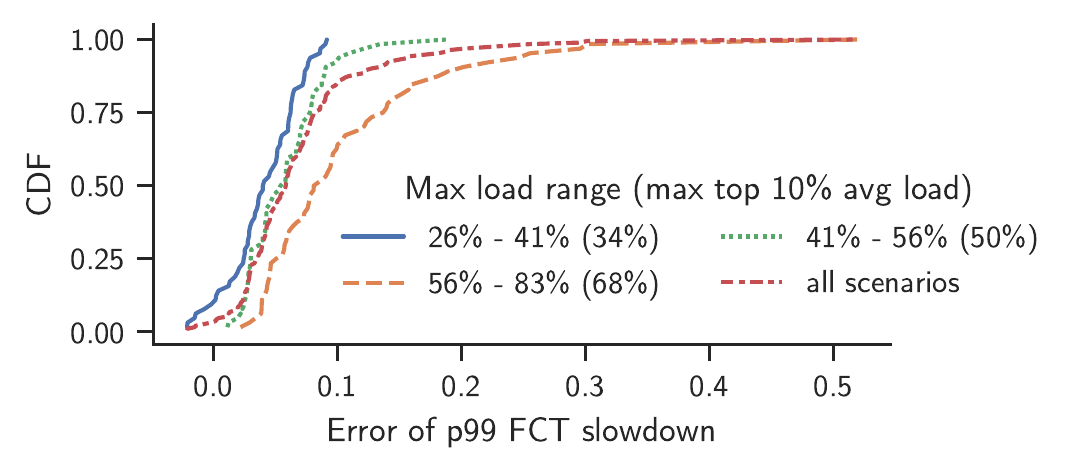}
    \vspace{-6mm}
    \caption{\small
        CDFs of p99 error between \sys and ns-3 across all scenarios drawn from
        the sample space in \Tab{sample-space}.
        The distributions are binned by maximum load.
        In parentheses, we give the maximum value for the top 10\% average load
        in each bin.
        Under common conditions of low to moderate load, \sys's estimates for
        the p99 FCT slowdown are reliably within 10\% of the ground truth.
    }
    \label{fig:error-dist-0}
\end{figure}

\begin{figure*}[t]
    \centering
    \begin{subfigure}[t]{0.45\linewidth}
        \centering
        \includegraphics[width=\textwidth]{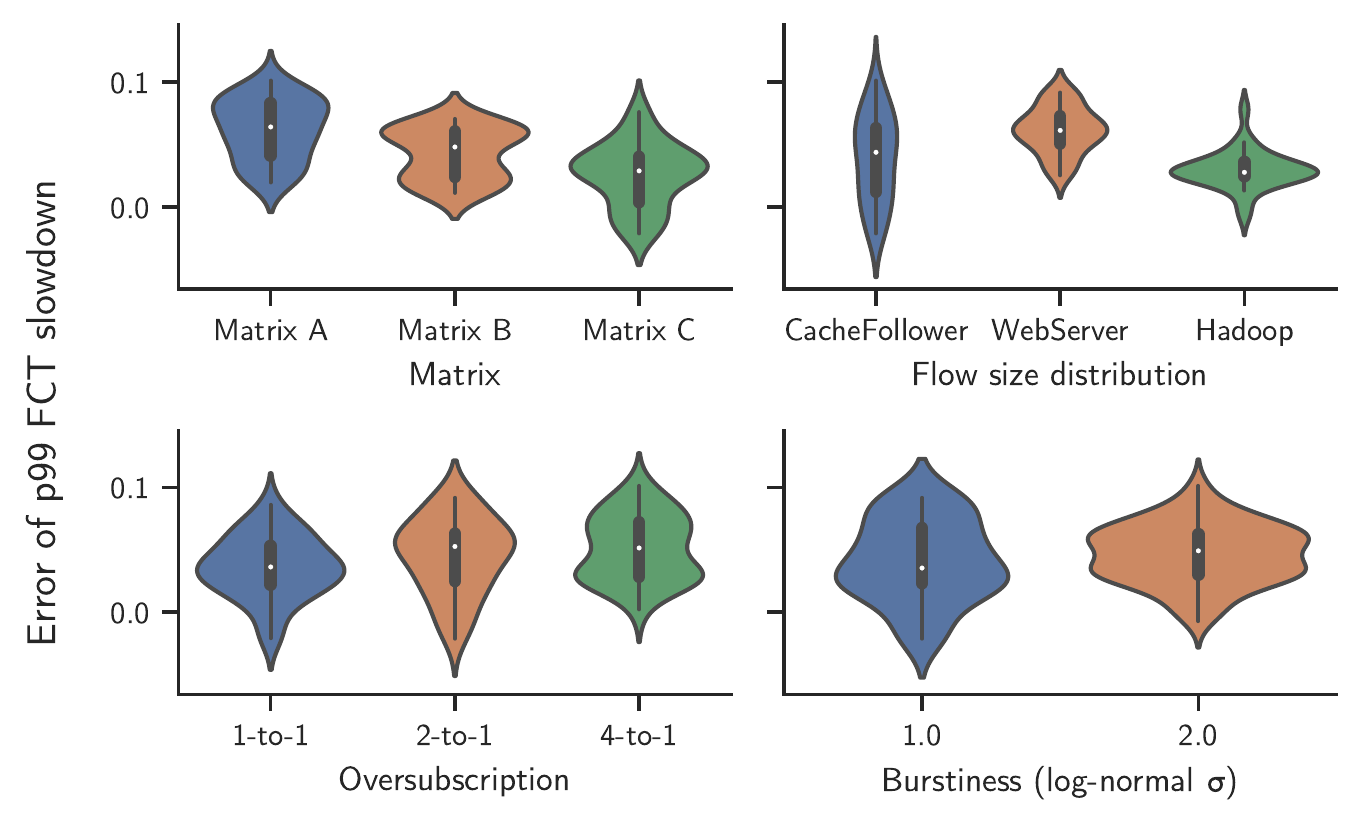}
        \vspace{-5mm}
        \caption{\small
            Max load $\leq$ 50\%
        }
        \label{fig:error-dist-1}
    \end{subfigure}
    \hfill
    \begin{subfigure}[t]{0.45\linewidth}
        \centering
        \includegraphics[width=\textwidth]{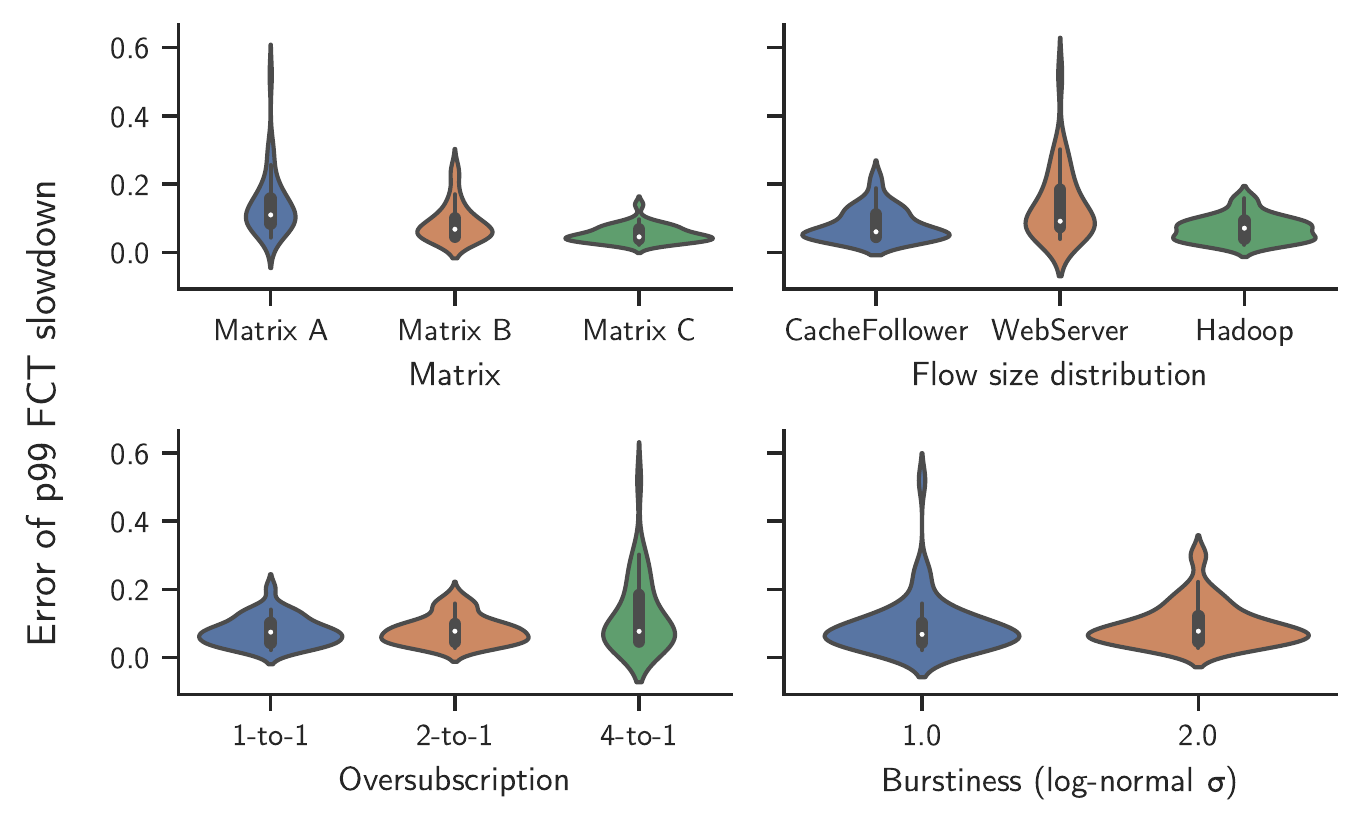}
        \vspace{-5mm}
        \caption{\small
            Max load $>$ 50\%
        }
        \label{fig:error-dist-2}
    \end{subfigure}
    \vspace{-3mm}
    \caption{\small
        Distributions of p99 error between \sys and ns-3, faceted by different
        workload and topology parameters.
        For each distribution we show the median, the quartiles, and a rotated
        kernel density estimation.
        We consider the low-load regime (\Fig{error-dist-1}) and the high-load
        regime (\Fig{error-dist-2}) separately.
        At low load, the workload and topology parameters only have a modest
        effect on \sys's accuracy, but at high load, the conditions leading to
        the largest errors come into view: high load, high oversubscription,
        with very short flows.
        Note the different y-axes between the two load regimes.
    }
    \label{fig:other-variables}
    \vspace{-2mm}
\end{figure*}

Next we turn our attention to how different aspects of workloads and
topologies affect \sys's accuracy.
To be able to simulate enough scenarios in ns-3 for a sensitivity analysis, we
downsample the topologies and traffic matrices to 32 racks.
The resulting topologies have two pods, 16 racks per pod, and eight hosts per
rack, and the number of spines per plane varies to accommodate different
oversubscription factors.

Our approach is as follows.
First, we construct a sample space over the parameters defining the workload
and the topology (aside from the number of servers, which is fixed).
%
The sample space is shown in \Tab{sample-space}.
Then, we sample 192 scenarios uniformly at random, and we run ns-3 and the
default \sys variant on each of them for several seconds of simulated time.
Next, for each scenario, we take the p99 FCT slowdown estimated by both ns-3
and \sys, and we compute the error between them.
If these values are $n$ and $p$ respectively, then the error is $(p - n) / n$.
Negative values indicate that \sys produced an underestimate.

Since we have one error value per scenario, the errors give rise to
distributions of error associated with the original sample space.
Now what remains is to determine how the workload and topology parameters
affect error distributions.
To start, recall from the discussion in \Sec{error-sources} that the magnitude
of error is expected to be load-dependent, with higher errors typically
manifesting at higher loads, so we begin by examining the effect of the maximum
load setting on \sys's accuracy.

\Para{Maximum load.} \Fig{error-dist-0} shows the error distributions binned by
maximum load.
Among all scenarios, \sys's p99 estimates are within 10\% of ns-3's estimates
85\% of the time
At high load, we observe larger overestimates of up to 52\% in the worst case.
In the most highly-loaded group of scenarios---with maximum link loads
between 56\% and 83\%---\sys is within 10\% of ns-3 62\% of the time, with an
average error of about 11\%.
%
However, this includes scenarios where 10\% of the links have an average load
of up to 68\%, which is much higher than what is reported in the literature.
For example, Roy \etal report that in Meta's data center network, 99\% of host
links are less than 10\% loaded, and the top 5\% of core links have loads
between 23\% and 46\%~\cite{fb-network}.
Among scenarios where the maximum link load is between 26\% and 41\%, \sys is
within 10\% of ns-3 100\% of the time.
If we further include scenarios with maximum link loads between 41\% and 56\%,
that number falls to 96\%.
Finally, while \sys's techniques tend to overestimate latencies, in
3\% of the scenarios, \sys underestimates p99 slowdown by up to 2\%.

\begin{table}
    \small
    \centering
    \begin{tabular}{rrcllr}
        \toprule
        Error  & Max load & Matrix & Sizes     & Oversub & $\sigma$ \\
        \midrule
        51.9\% & 77.6\%   & A      & WebServer & 4-to-1  & 1        \\
        30.1\% & 67.3\%   & A      & WebServer & 4-to-1  & 2        \\
        29.6\% & 67.0\%   & A      & WebServer & 4-to-1  & 2        \\
        25.6\% & 65.9\%   & A      & WebServer & 4-to-1  & 1        \\
        24.6\% & 73.2\%   & B      & WebServer & 4-to-1  & 1        \\
        \bottomrule
    \end{tabular}
    \caption{\small
        The five scenarios with the highest error values from the sensitivity
        analysis in \Sec{sensitivity}.
    }
    \label{tab:top5-error}
    \vspace{-6mm}
\end{table}

\Para{Other parameters.} We next turn to the effects of all other workload and
topology parameters.
We start by only considering scenarios where the maximum link load is less than
or equal to 50\%; this will tell us whether any of the parameters have a large
effect on accuracy in the low-load regime.
\Fig{error-dist-1} shows the median error and error distributions as a violin plot for low-load scenarios grouped
by traffic matrix, flow size distribution, oversubscription, and burstiness.
Overall, changes to these parameters appear only to have a modest effect.
The choice of traffic matrix has the clearest trend, but load is a confounder
here: recall from \Fig{link-loads} that different traffic matrices yield
different link load distributions for the same maximum load setting.

When we look at the high load regime in \Fig{error-dist-2}, a clear picture
comes into view.
We see much longer tails in error distributions for matrix A, the WebServer
flow size distribution, and 4-to-1 oversubscription.
Together with \Fig{error-dist-1}, this suggests that none of these settings has
a strong effect on its own, but \emph{coupled together in the high load
    regime,} they have a pronounced effect on \sys's accuracy.
Matrix A induces higher average load and has more cross-rack traffic, making it
more likely for its flows to encounter multiple simultaneous bottlenecks.
The WebServer flow size distribution is dominated by short
flows (\Fig{flow-sizes}), a third of which are smaller than 1 KB and 80\% of
which are smaller than 10 KB. Because more of the traffic completes within
a single round trip, there is more ephemeral congestion and bandwidth smoothing
can have a larger impact.
%

Finally, oversubscription has an effect at high load: if
we removed all scenarios with 4-to-1 oversubscription, the maximum error would
only be 20\% rather than 52\%, even at high load.
In addition to the double counting of delays described in \Sec{error-sources},
oversubscription can also increase correlations in link delays.
To achieve 4-to-1 oversubscription in topologies as small as these, there are
only four spine switches per plane forwarding traffic between groups of 16
racks, leaving relatively few paths through the core.
Fewer paths can result in higher degrees of correlation---especially with matrix
A, whose traffic is primarily inter-rack (\Fig{matrices}).
Finally, this setting combined with the short flows from the WebServer
distributions gives rise to errors of up to 52\%.

\Tab{top5-error} lists the scenarios with the top five highest error values.
Four have matrix A, all have the WebServer distribution, and all five have
4-to-1 oversubscription.
In this group, the average maximum load is 70.2\%.
\revv{
    In practice, we believe it is rare to find all-to-all workloads on heavily
    oversubscribed topologies where core links are persistently 70\% utilized.
    Consequently,
}{
    Since we expect the combination of all-to-all workload, heavily
    oversubscribed topology, and persistently high core utilization to occur
    relatively infrequently,
}
the data suggest that \sys maintains good accuracy under common
conditions.

\rev{\Para{Mixed Workloads.} We also use the small topology to study the \sys prediction error for
subsets of traffic in heterogeneous workloads in Appendix \Sec{workload-mixes}.}

\begin{figure}[t]
    \centering
    \includegraphics[width=\linewidth]{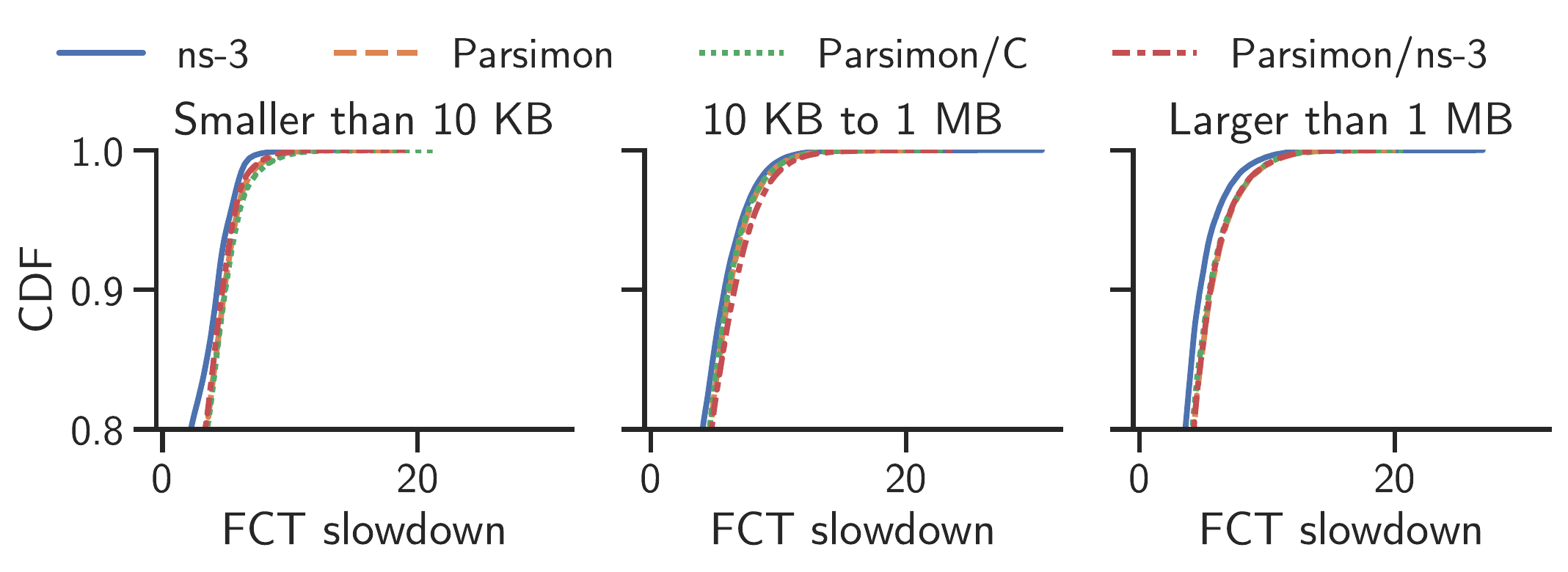}
    \vspace{-6mm}
    \caption{\small
        CDFs of FCT slowdown estimated by ns-3 and \sys for the scenario whose
        error is at the 85\textsuperscript{th} percentile of the p99 error
        distribution.
        Note the y-axis.
        Even though the accuracy here is worse than in the common case, \sys's
        estimates remain close across most of the tail.
        Also shown is \sys/ns-3.
    }
    \label{fig:one-scenario}
    \vspace{-3mm}
\end{figure}

\revv{\Para{Analyzing one scenario.} Finally, we pick one scenario to analyze so that we
    can see the error of the entire tail rather than just the p99.
    To pick one whose accuracy is worse than the average case, we rank-order all
    scenarios by error and select the one at the 85\textsuperscript{th} percentile.
    We find that it has matrix A, the Hadoop flow size distribution, low
    burstiness, 2-to-1 oversubscription, and a maximum load of 68\% (with a top
    10\% average load of 56\%).
}{}

\revv{
    \Fig{one-scenario} shows that the accuracy at the p99 generalizes to other
    percentiles in the tail.
    We also run the scenario in \sys/C and \sys/ns-3 (\Tab{variants}), and we find
    little accuracy difference between any of the variants.
    For 14 seconds of simulated time, ns-3 took nearly five hours, \sys/ns-3
    took 34 minutes, \sys took 1.5 minutes, and \sys/C took 49 seconds.
    The speed-up of \sys/ns-3 being modest (see \Sec{fast-sim} for discussion), it
    may appear at first glance as though most of \sys's benefit comes from the
    custom backend.
    However, gains due to the custom backend are fixed at an order of magnitude,
    while gains due to decomposition scale with the number of cores and the size of
    the simulated network.
    In addition, decomposition is what enables other optimizations like clustering.
}{}

\rev{
    \subsection{Analysis of One Configuration} \label{s:oneconfig}
}
We pick one representative scenario to examine in more detail, to test
if our approach is robust to alternate definitions of tail latency, congestion control protocol,
workload, and topology.
To pick a scenario whose accuracy is somewhat worse than the average case, we rank-order all
scenarios by error and select the one at the 85\textsuperscript{th} percentile.
This has matrix A, the Hadoop flow size distribution, low
burstiness, 2-to-1 oversubscription, and a maximum load of 68\% (with a top
10\% average load of 56\%).

\Para{Tail distribution.}
Operators may differ in their definitions of tail latency, e.g., focusing on the 90th
or 99.9th percentile, rather than just the 99th FCT slowdown.  \Fig{one-scenario} shows the
tail of the cumulative distribution of FCT slowdown for different flow sizes for the selected configuration,
for ns-3 and each of the \sys variants.  The prediction error is similar across the tail of the distribution
for this scenario, with little accuracy difference between any of the variants.

\begin{table}
    \small
    \centering
    \begin{tabular}{lcccc}
        \toprule
        Protocol & Max load & \multicolumn{3}{c}{Error in p99 FCT slowdown}                         \\
                 &          & < 10 KB                                    & 10 KB - 1 MB & > 1 MB \\
        \midrule
        DCTCP    & 45\%     & 1.4\%                                      & 9.2\%        & 15.9\% \\
        TIMELY   & 45\%     & 4.0\%                                      & 17.9\%       & 13.7\% \\
        DCQCN    & 45\%     & 5.9\%                                      & 11.6\%       & 12.8\% \\
        \midrule
        DCTCP    & 56\%     & 2.8\%                                      & 9.2\%        & 14.6\% \\
        TIMELY   & 56\%     & 8.1\%                                      & 20.0\%       & 11.3\% \\
        DCQCN    & 56\%     & 7.6\%                                      & 14.6\%       & 12.2\% \\
        \midrule
        DCTCP    & 67\%     & 13.8\%                                     & 11.3\%       & 13.6\% \\
        TIMELY   & 67\%     & 13.3\%                                     & 18.2\%       & 5.0\%  \\
        DCQCN    & 67\%     & 18.0\%                                     & 15.2\%       & 13.6\% \\
        \bottomrule
    \end{tabular}
    \caption{\small
        Prediction error of \sys/ns-3 for estimated p99 FCT slowdown with three
        different congestion control protocols for the sample configuration at different load levels
        and for different request sizes.
    }
    \label{tab:cc-kinds}
\end{table}

\Para{Transport protocols.}
We use the sample scenario to test the generality of \sys to two additional
congestion control protocols, DCQCN~\cite{dcqcn} and TIMELY~\cite{timely}.
DCQCN is designed for RDMA traffic, while TIMELY uses network delay,
rather than ECN signals, to detect congestion.
To focus on prediction error for our approximation methods, we use the pre-existing ns-3 implementation
of the protocols as the \sys link level simulator for this
part of the evaluation.  Note that \sys and \sys/ns-3 exhibit a few percent
difference in p99 error for DCTCP for this configuration.  Because the
prediction error for different congestion control protocols may depend
on the amount of congestion, we also run the experiment at varying load levels.

\Tab{cc-kinds} shows the prediction error for \sys/ns-3 relative to ns-3 in the estimated p99 FCT
slowdown at three load levels for the three transport protocols, aggregated by request size.
For this configuration, \sys is most accurate for small flows and low to moderate maximum link
utilization, and that is true for all three congestion control protocols.  DCTCP has somewhat
lower error for small and medium size flows at low to moderate utilization.
Relative error is higher for larger transfers and higher maximum link utilization,
with no clear pattern in the error for different protocols.

\Para{Simulated link failures.}
We also use the sample configuration to examine the prediction accuracy for
topologies with simulated link failures in Appendix \Sec{link-failures}.

%% file: conclusion.tex
\section{Conclusion}
In this paper, we propose and evaluate a new method
for computing a conservative estimate of flow-level tail latency
for large scale data center networks, given an arbitrary
traffic matrix, topology, flow size distribution, and inter-arrival
process. Our approach decomposes the problem into a large number of
individual link simulations, specially constructed
to produce accurate estimates of the probability distribution of
delay contributed by congestion at each link.  We then mechanically combine these
link-level delay distributions to produce flow-level estimates.
On a large-scale network using a commercial workload,
our approach outperforms ns-3 by a factor of 492
on a single multicore server with a loss of accuracy of
less than 9\% in the tail of the latency distribution.

\smallskip
\noindent
{\color{black}\rev{\textbf{Acknowledgments.} We are grateful to Vincent Liu, Jeff Mogul, our shepherd Arpit Gupta, and the anonymous reviewers for their feedback and useful comments. This work was supported in part by NSF grants CNS-2006346, CNS-2006827, a Cisco Research Center Award, and a Google Research Award.}}

%% file: appendix.tex
\input{appendix/eval-ext.tex}
\input{appendix/parking-lot.tex}
\input{appendix/clustering.tex}

%% file: appendix/eval-ext.tex
\rev{
\section{Mixed Workloads} \label{s:workload-mixes}   
}


\sys's methods are designed to estimate performance
distributions rather than per-flow metrics.
However, it is often useful to aggregate FCT performance estimates in different
ways.
For example, an operator may wish to estimate the performance of individual
virtual networks or individual services.
In this section, we conduct a simple experiment to assess \sys's ability to
estimate performance for separate aggregates.

\begin{figure}[t]
    \centering
    \includegraphics[width=\linewidth]{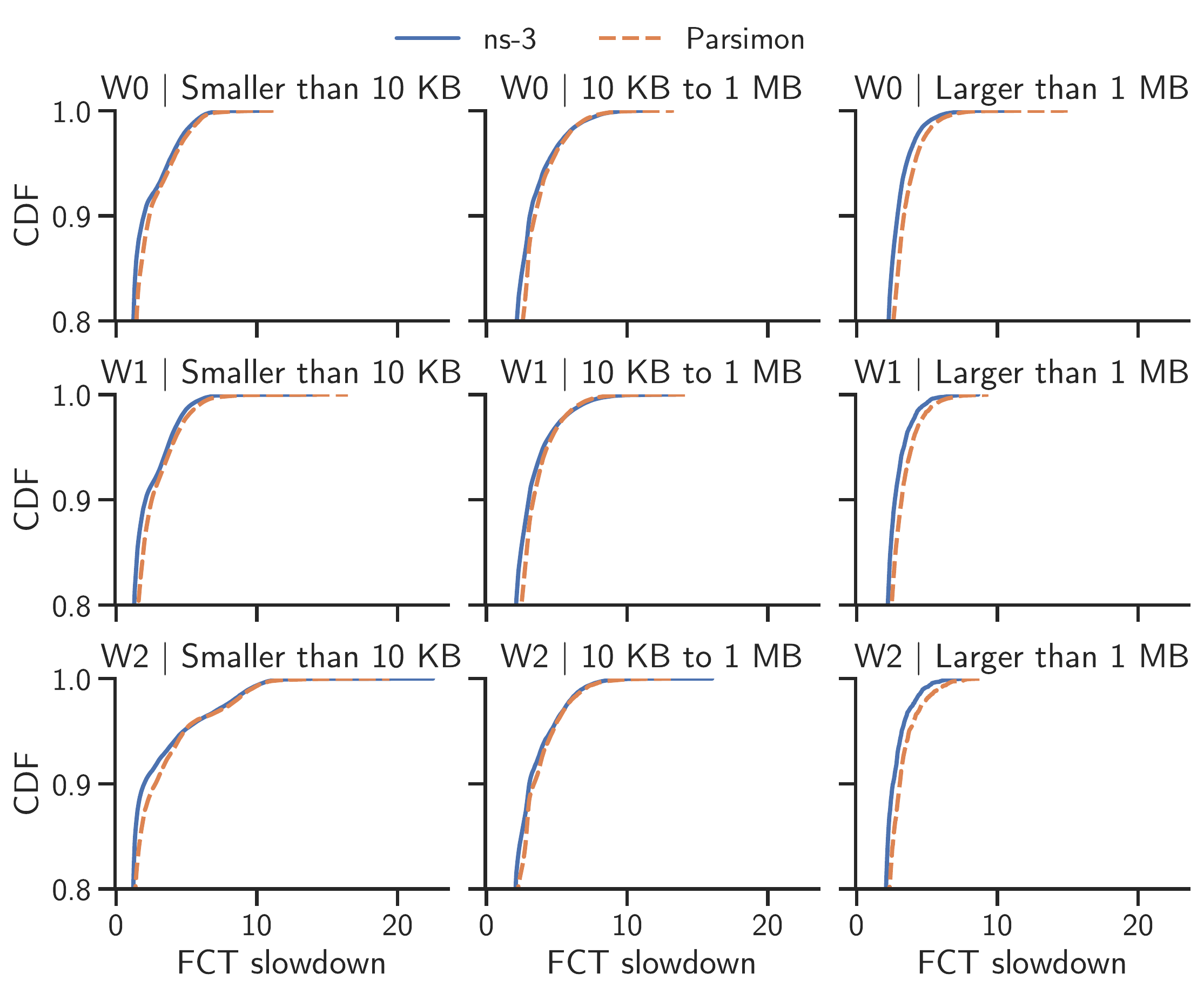}
    \caption{\small
        CDFs of FCT slowdown for ns-3 and \sys, bucketed by workload and flow
        size.
        Note the y-axes.
        When mixing workloads in a single simulation, \sys can accurately
        estimate performance distributions for individual workloads in addition
        to full-network aggregates.
    }
    \label{fig:workload-mixes}
\end{figure}

We start by mixing three different workloads---each with its own traffic matrix and flow size distribution---into
one workload.
\Tab{workload-mixes} summarizes their differences.
Each workload has a maximum load setting of 20\% and a high burstiness setting
($\sigma = 2$), and their combination results in a maximum link load of about
50\%.
We run the combined workload on the small-scale topology with 2-to-1
oversubscription from \Sec{sensitivity}, and we observe the accuracy for each
workload faceted by flow size.
\Fig{workload-mixes} shows the cumulative distribution function (CDF) of FCT slowdown 
for ns-3 and \sys.
We observe that across all workloads and flow size bins, \sys maintains good
accuracy.

\begin{table}
    \small
    \centering
    \begin{tabular}{lclrr}
        \toprule
        Name & Matrix & Sizes         & Max load            & $\sigma$ \\
        \midrule
        W0   & A      & CacheFollower & \textasciitilde20\% & 2        \\
        W1   & B      & WebServer     & \textasciitilde20\% & 2        \\
        W2   & C      & Hadoop        & \textasciitilde20\% & 2        \\
        \bottomrule
    \end{tabular}
    \caption{\small
        The three workloads mixed together in \Sec{workload-mixes}.
    }
    \label{tab:workload-mixes}
\end{table}

\rev{
\section{Link Failures} \label{s:link-failures}
}

\begin{figure}[t]
    \centering
    \begin{subfigure}[t]{0.35\linewidth}
        \centering
        \includegraphics[width=\textwidth]{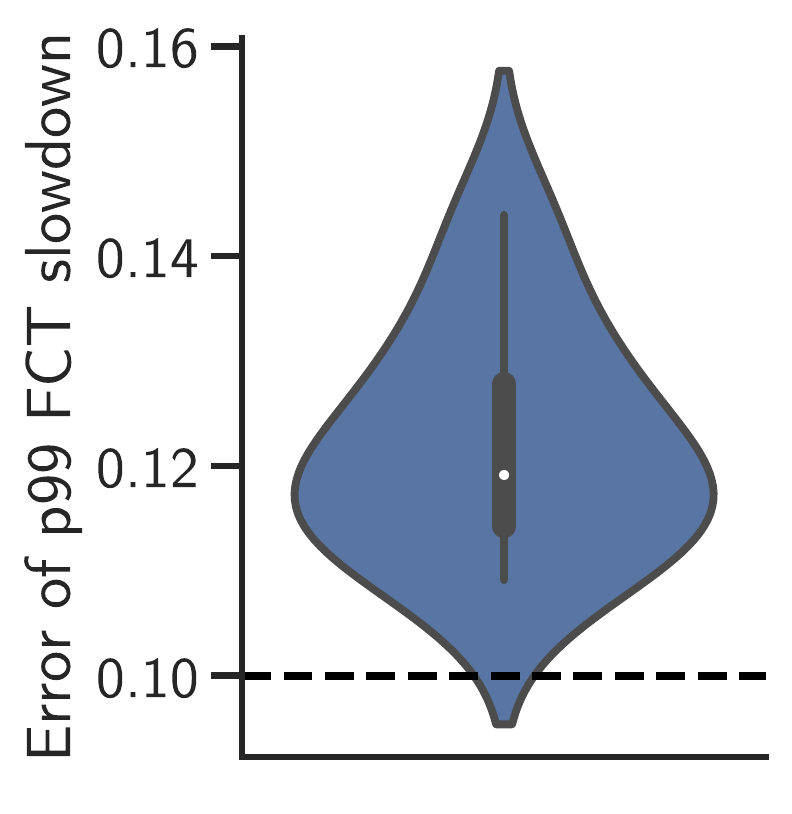}
        \caption{\small
            p99 errors
        }
        \label{fig:link-failures-p99}
    \end{subfigure}
    \hfill
    \begin{subfigure}[t]{0.63\linewidth}
        \centering
        \includegraphics[width=\textwidth]{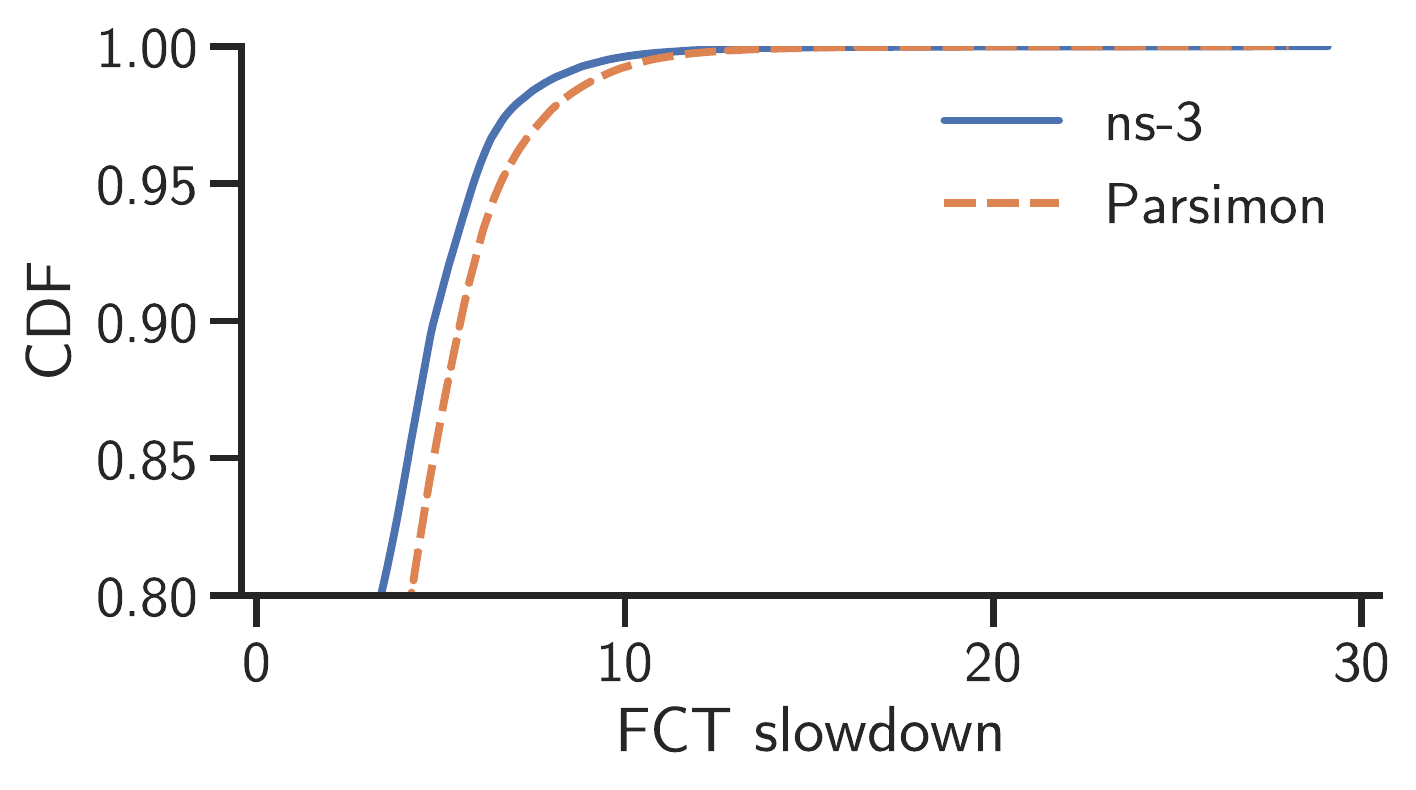}
        \caption{\small
            CDF with the max p99 error (0.144)
        }
        \label{fig:link-failures-9}
    \end{subfigure}
    \caption{\small
        Errors between ns-3 and \sys in estimated FCT slowdowns when there is a
        link failure.
        \Fig{link-failures-p99} shows the error distribution for p99 estimates
        from ten trials---each with one random link failure---with the dashed
        line showing the error with no link failure.
        \Fig{link-failures-9} shows
        the CDF of FCT slowdowns for the trial with the highest p99 error.
        For the small oversubscribed topology used in this experiment, a link
        failure modestly increases estimation error.
    }
    \label{fig:link-failures}
\end{figure}

One operational use case for \sys is to estimate counter-factual network performance in the
presence of potential link failures or planned outages. In this section, we use the sample scenario
from \Sec{oneconfig} (matrix A, the Hadoop flow size distribution, low burstiness, 2-to-1
oversubscription, and a maximum link load of 68\%) to evaluate \sys for this use case. 
For this configuration, the error in estimated p99 FCT slowdown between ns-3 and \sys was around 10\%.
Since link failures increase the load on the remaining links, we should expect some decreased
accuracy for \sys in this case.  On the other hand, simulating all possible network failures in ns-3
would be prohibitively expensive.

In selecting links to fail, we only consider links in ECMP groupings, such that
the failure of one link causes traffic to be routed to the other links in the
group.
In Meta's data center fabric~\cite{fb-fabric}, this corresponds to links
between fabric switches and spine switches and links between ToR switches and
fabric switches.
In the small 32-rack topology used here (\Sec{sensitivity} for details), there
are 96 such links.
We run ten trials, each time picking a random one of the
links to fail, keeping the workload constant.
We note that this setting represents a particularly bad case for \sys: in addition to
the high link loads, the scenario uses an all-to-all communication pattern on a small and
oversubscribed topology, which means each link failure in the core can have an
outsized effect on other core links.

\Fig{link-failures-p99} shows the distribution of errors in p99 estimates.
%
With a single link failure, the errors range from 11\% to 14\%, with a median error of
12\%.
\Fig{link-failures-9} shows the estimated CDFs of FCT slowdown for the trial
with the highest error.

%% file: appendix/parking-lot.tex
\rev{
    \section{Studying Error Sources} \label{s:app-error-sources}
}

\begin{figure}[t]
    \centering
    \includegraphics[width=0.9\linewidth]{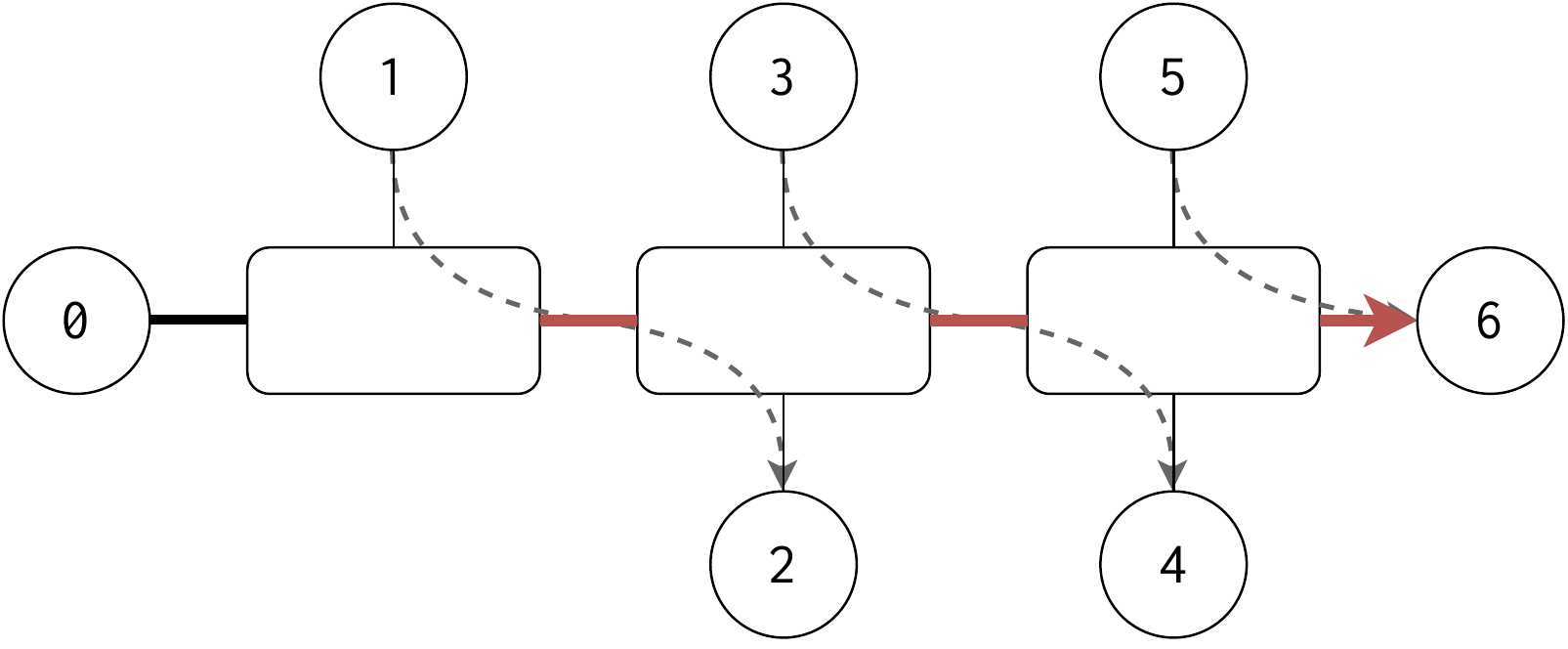}
    \vspace{-2mm}
    \caption{\small
        The parking lot topology used in \Sec{app-error-sources}.
        In this topology, zero sends to six, one sends to two, three sends to
        four, and five sends to six.
        We refer to the traffic from zero to six as \emph{main traffic} and to
        all other traffic as \emph{cross traffic}.
        The bolded red links contain both main traffic and cross traffic, and
        we call them \emph{congested links}.
    }
    \label{fig:parking-lot}
    \vspace{-4mm}
\end{figure}

Recall from \Sec{error-sources} that \sys's approximations induce errors in
its end-to-end estimates.
In this appendix, we use microbenchmarks to study the effects of some
pathological cases on \sys's accuracy.
For an initial discussion on these topics, please refer to \Sec{error-sources}.

Throughout, we use the parking lot topology shown in \Fig{parking-lot} with 40
Gbps links.
The flow of traffic through the topology is shown with arrows and described in
the caption.
We refer to the traffic from node zero to node six as \emph{main traffic} and
to all other traffic as \emph{cross traffic}.
The bolded red links contain both main traffic and cross traffic, and we call
them \emph{congested links}.
In all experiments, we set the load of the main traffic to 25\%.
When there is cross traffic, its load is also 25\%, yielding a total load of
50\% on all three congested links.
Lastly, to isolate the effects on the main path from zero to six, we measure
FCT slowdown distributions only for the main traffic.

\subsection{First-Hop Delays} \label{s:first-hop-delays}

\begin{figure}[t]
    \centering
    \includegraphics[width=\linewidth]{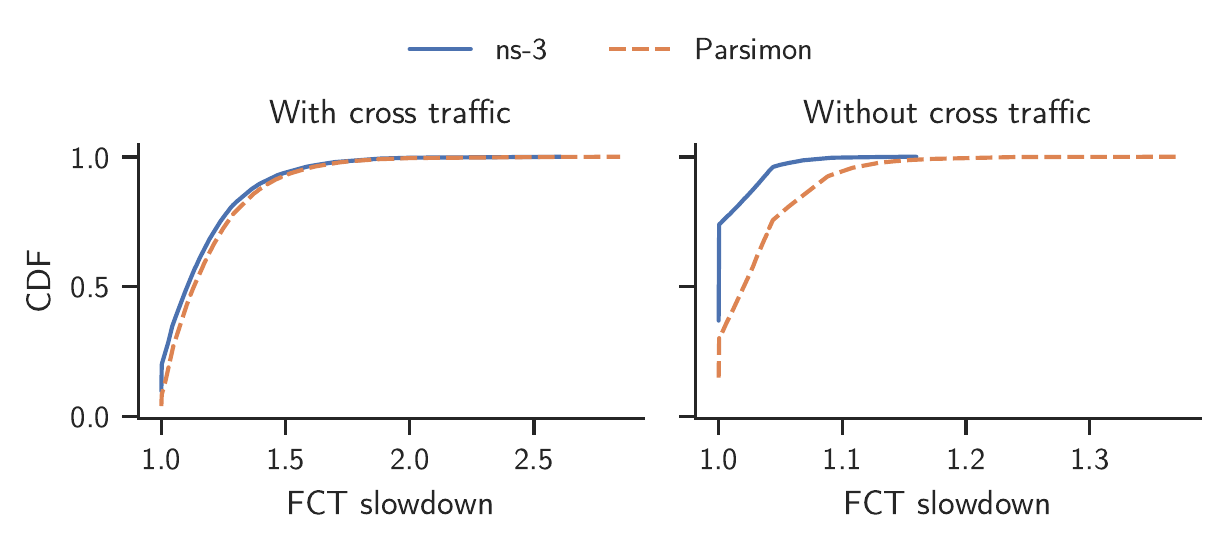}
    \vspace{-8mm}
    \caption{\small
        CDFs of FCT slowdown estimated by ns-3 and \sys for the main traffic,
        both with and without cross traffic.
        When there is cross traffic, errors arising from first-hop delays are
        second-order, as most delays are cause by queueing on the congested
        links.
        However, when there is no cross traffic, those errors become dominant.
        The graph on the right uses the same workload as the one on the left,
        except the cross traffic is removed.
        Note the different x-axes.
    }
    \label{fig:first-hop-delays}
    \vspace{-4mm}
\end{figure}

First, consider the case where all traffic in \Fig{parking-lot} originates from
node zero and is destined to node six, and recall that all links have the same
capacity.
In a real network, all queueing in this scenario would occur at the first hop.
Subsequent hops would see traffic completely smoothed, and they would therefore
contributing zero queueing delay.

If we re-examine how link-level topologies are constructed in \Fig{link-level},
we see that this smoothing effect is captured, since all traffic passes through
edge links with the original edge-link capacities.
However, for the link-level topologies in cases B and C of \Fig{link-level}, it
is possible for first-hop edge links to contribute delays that will be
(erroniously) attributed to the target link.
In most cases, we expect the magnitude of this error to be small.
A target link will almost always have multiple sources, and only the traffic
passing through the target link is simulated.
Consequently, the first-hop delays in link-level simulation are expected to be
small compared to delays accrued at target links.

The scenario which we first described---in which all traffic on a path
originates from a single source---represents the worst case.
Here, all target links (aside from the first hop) contribute no queueing delay,
thus magnifying the error induced by repeatedly counting the first-hop delays
for each target link.
\Fig{first-hop-delays} shows this effect.
In this experiment, the main traffic consists of one kilobyte flows, and the
cross traffic consists of 10 kilobyte flows.
All traffic follows a Poisson arrival process.
With cross traffic, we see from the graph on the left that \sys accurately
estimates the FCT slowdown distribution of the main traffic.
However, when we remove the cross traffic, as done to produce the graph on the
right, we see substantial error in \sys's estimates due to the first-hop delays
previously described.
We note that this error exists \emph{even when there is cross traffic}, but the
error contributes so little to total delays---which are dominated by queueing
at congested links---that \sys still maintains good accuracy.

\subsection{Correlated and Simultaneous Delays} \label{s:correlations}

\begin{figure}[t]
    \centering
    \begin{subfigure}[t]{\linewidth}
        \centering
        \includegraphics[width=\textwidth]{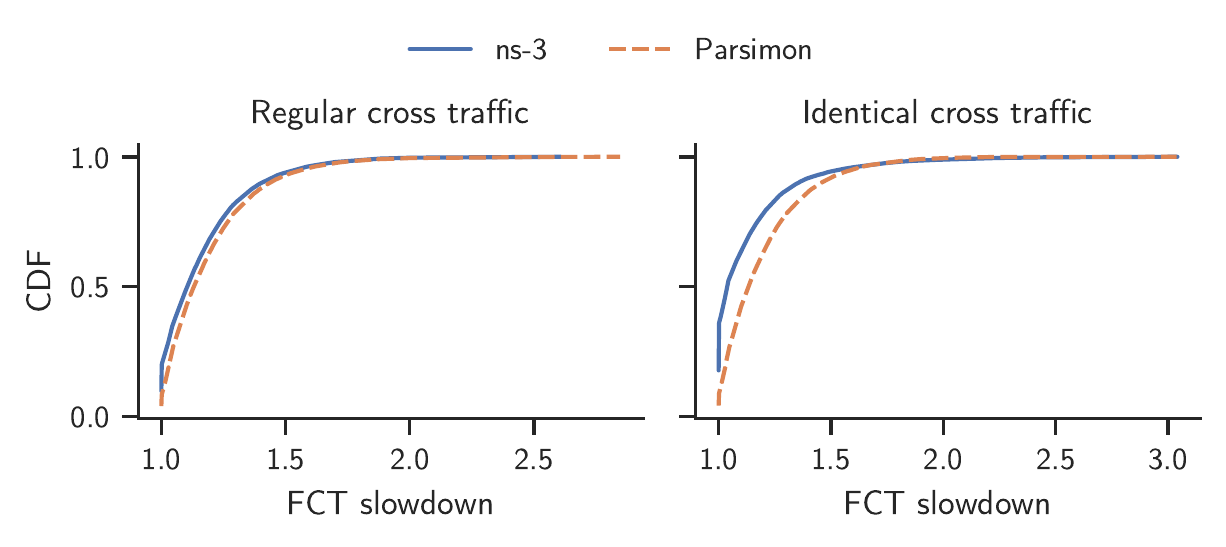}
        \vspace{-4mm}
        \caption{\small
            Short flows (1 KB), Poisson cross traffic
        }
        \label{fig:short-corr}
    \end{subfigure}
    \hfill
    \begin{subfigure}[t]{\linewidth}
        \centering
        \includegraphics[width=\textwidth]{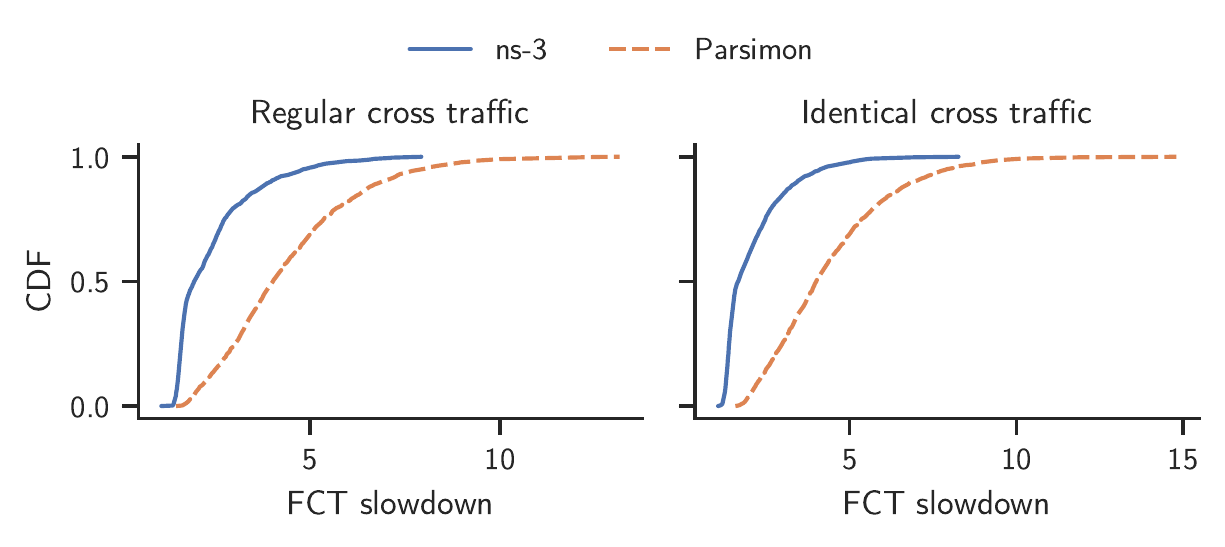}
        \vspace{-4mm}
        \caption{\small
            Long flows (400 KB), Poisson cross traffic
        }
        \vspace{-1mm}
        \label{fig:long-corr}
    \end{subfigure}
    \caption{\small
        CDFs of FCT slowdown estimated by ns-3 and \sys for the main traffic
        with regular or identical cross traffic.
        The main traffic consists either of short flows (\Fig{short-corr}) or
        long flows (\Fig{long-corr}).
        When delays are artifically correlated by replicating the same cross
        traffic across hosts, accuracy decreases for both short and long flows,
        with long flows seeing larger errors.
        In fact, long-flow estimates have significant error even when delays
        are not explicitly correlated; this is due to the simultaneous delays
        induced by the smooth Poisson cross traffic.
    }
    \label{fig:corr-poisson}
    \vspace{-2mm}
\end{figure}

\begin{figure}[t]
    \centering
    \includegraphics[width=\linewidth]{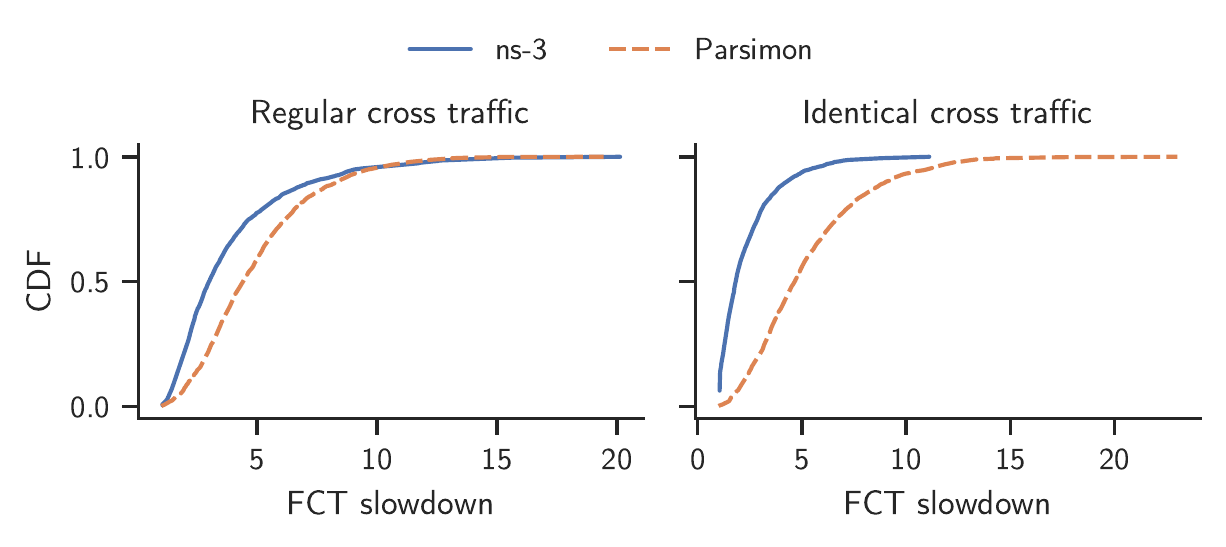}
    \vspace{-8mm}
    \caption{\small
        CDFs of FCT slowdown for the same scenario as in \Fig{long-corr}, but
        with bursty cross traffic (log-normal inter-arrival times, $\sigma=2$).
        When the cross traffic is bursty, long flows experience fewer
        simultaneous delays with regular cross traffic.
        This results in less error in \sys's estimates.
    }
    \label{fig:long-corr-2}
    \vspace{-4mm}
\end{figure}

Next we examine the effect of correlated and simultaneous delays on \sys's
accuracy.
We begin by artificially correlating delays and examining the effect on
estimated slowdown distributions.
Note that if the delays along a path are positively correlated---for example,
if the probability of encountering delay at hop $i+1$ is higher given there is
delay at hop $i$---then we also expect to see more simultaneous delays along
the path.
We create these correlated delays by modulating the cross traffic.
For regular unmodified cross traffic, we use the same setup as in the previous
subsection (\Sec{first-hop-delays}).
To artificially correlate delays, we replicate the exact sequence of flows from
source one on sources three and five in \Fig{parking-lot}, so that all three
sources of cross traffic send the same flows at the same time.
This produces an extreme case of correlation.

Because short-flow and long-flow estimates have different sources of error,
we separate the two cases when generating the main traffic.
For short flows we use the same one kilobyte flows as before, and for long
flows we generate flows that are 10 times the maximum bandwidth-delay product, or 400 kilobytes.
\Fig{corr-poisson} shows the effect of correlating delays on \sys's accuracy
for short and long flows.

\Para{Short-flow main traffic.}
In the case of short flows (\Fig{short-corr}), a chief effect of increased
correlation is to alter the probability that a flow will encounter queueing.
For example, suppose a short flow traverses only two links at 50\% utilization.
If the delays of the two links are independent, we can estimate the probability
that the flow encounters no delay (\ie no queueing) as $50\% \times 50\% = 25\%$.
However, if the delays are perfectly positively correlated, then the
probability that the flow encounters no delay increases to 50\%.
\sys does not capture this effect because it treats all links independently; in
this experiment, this manifests as slight overestimates in FCT slowdown
distributions.

\Para{Long-flow main traffic.}
While the total delay for a short flow can be thought of as the sum of
individual link delays, the same reasoning does not straightforwardly extend to
long flows.
Unlike a short flow, a long flow occupies multiple hops at the same time, and
only the bottleneck at each instant contributes to end-to-end delay.
Summing link delays is therefore only appropriate if different hops contribute
significant delays at largely different times.
However, \sys always aggregates individual link contributions by adding them,
regardless of whether a link was the bottleneck when the delay was incurred.
When we turn our attention to \Fig{long-corr}, we see that not only is the
effect of identical cross traffic more severe, but also there is significant
error even with regular cross traffic.
This is because the cross traffic is smooth (recall that it uses uniform flow
sizes and a Poisson arrival process).
Smooth traffic results in small but frequent delays at congested links,
increasing the chance that long flows will experience simultaneous delays.

In \Fig{long-corr-2}, we duplicate the scenario in \Fig{long-corr}, except we
make the cross traffic bursty by using a log-normal inter-arrival time
distribution with shape parameter $\sigma=2$.
Because the cross traffic is bursty, there is less simultaneous delay in the
regular case, and the induced error is less dominant.
Consequently, \sys's estimates are closer to the ground truth in the graph on
the left.
Identical cross traffic still induces large and frequent simultaneous delays,
so the errors remain in the graph on the right.

%% file: appendix/clustering.tex
\section{Clustering Details} \label{s:more-clustering}

Here we briefly describe the distance function and the thresholding critera we
use in the evaluation (\Sec{eval}) for clustering link-level simulations.
First, recall from \Sec{clustering} that the link features we extract are 1)
the average load, 2) the flow size distribution, 3) the inter-arrival time
distribution.
For any two links, we compute distances between their features, and we cluster
the links together if the distances are under some threshold.

\Para{Distance functions.} To compute a distance between link loads, we compute
the error.
If $a$ and $b$ are two link loads, error $e$ is computed as
\begin{equation*}
    e = \frac{|a - b|}{a}.
\end{equation*}
To compare distributions, there are many options.
We opt for a function that is 1) easily interpretable, 2) scale-independent,
and 3) adequately captures differences in the tail.
To compute a distance between two distributions, we extract 1,000
percentiles from each of them, and we compute a weighted mean absolute
percentage error (WMAPE) between them.
Suppose $A$ and $B$ are the sequences of extracted percentiles.
Then, WMAPE is computed as
\begin{equation*}
    \text{WMAPE} \ = \ \frac{
    \sum_{i=1}^{n} |A_{i} - B_{i}|
    }{
    \sum_{i=1}^{n} |A_{i}|
    }.
\end{equation*}

\noindent For our purpose, $A_{i}$ and $B_{i}$ are non-negative for all $i$.
We note it is a bit counterintuitive for our distance functions not to commute.
However, we have found that it is easy to set thresholds for these metrics, and
they produce adequate clustering for the workloads under study.

\Para{Distance thresholds.} Recall that we only want to cluster two links
together if we expect their simulation outputs to be similar.
Consequently, when setting a threshold for link loads we must consider the
network and the workload being assessed.
At high load, small differences in link loads can yield large differences in
the tails of FCT distributions; in these cases, we typically set tighter
thresholds to preserve accuracy (as usual, this is subject to a speed-accuracy
trade-off).
For highly-loaded networks, we commonly require $e < 0.001$ or $e < 0.002$ for
links to be clustered together.
Ideally, this decision would be made on a link-by-link basis, so that tighter
thresholds would be set only for high-load links---doing so may allow for more
liberal clustering of the low-load links contributing little delay, resulting
in more pruned simulations.
However, the current prototype sets a single threshold per simulation.
To set a threshold between distributions, we typically require $\text{WMAPE} <
    0.1$.